\documentclass[sigconf]{acmart}
\AtBeginDocument{%
  }

\acmConference[SIGMOD '26]{ACM Special Interest Group on Management of Data}{May 31 - June 05, 2026}{Bengaluru, India}
\acmDOI{10.1145/3769844}

\usepackage{amsmath,amsfonts}
\usepackage{caption}
\usepackage{graphicx}
\usepackage{subcaption}
\usepackage{algorithm}
\usepackage{algpseudocode}
\usepackage{minted}
\usepackage{listings}
\usepackage{multirow}
\usepackage{tabularx}
\usepackage{array}
\usepackage{marginnote}

\begin{document}

\title{MorphingDB: A Task-Centric AI-Native DBMS for Model Management and Inference}



\author{Sai Wu}
\affiliation{%
  \institution{Zhejiang University}\country{China}
  }
\email{wusai@zju.edu.cn}

\author{Ruichen Xia}
\affiliation{%
  \institution{Zhejiang University}\country{China}
  }
\email{rcxia@zju.edu.cn}

\author{Dingyu Yang}
\affiliation{%
  \institution{Zhejiang University}\country{China}
  }
\email{yangdingyu@zju.edu.cn}

\author{Rui Wang}
\affiliation{%
  \institution{Zhejiang University}\country{China}
  }
\email{rwang21@zju.edu.cn}

\author{Huihang Lai}
\affiliation{%
  \institution{Institute of Computing Innovation, Zhejiang University}\country{China}
  }

\author{Jiarui Guan}
\author{Jiameng Bai}
\affiliation{%
  \institution{Zhejiang University}\country{China}
  }
  
\author{Dongxiang Zhang}
\author{Xiu Tang}
\affiliation{%
  \institution{Zhejiang University}\country{China}
  }
  
\author{Zhongle Xie}
\author{Peng~Lu}
\affiliation{%
  \institution{Zhejiang University}\country{China}
  }
\author{Gang Chen}
\email{cg@zju.edu.cn}
\affiliation{%
  \institution{Zhejiang University}\country{China}
  }



\renewcommand{\shortauthors}{Sai Wu et al.}

\newcommand{\morphDB}{MorphingDB}

\begin{abstract}
The increasing demand for deep neural inference within database environments has driven the emergence of AI‑native DBMSs. However, existing solutions either rely on model-centric designs requiring developers to manually select, configure, and maintain models, resulting in high development overhead, or adopt task-centric AutoML approaches with high computational costs and poor DBMS integration. We present MorphingDB, a task‑centric AI-native DBMS that automates model storage, selection, and inference within PostgreSQL. To enable flexible, I/O-efficient storage of deep learning models, we first introduce specialized schemas and multi-dimensional tensor data types to support BLOB‑based all‑in‑one and decoupled model storage. Then we design a transfer learning framework for model selection in two phases, which builds a transferability subspace via offline embedding of historical tasks and employs online projection through feature-aware mapping for real‑time tasks. To further optimize inference throughput, we propose pre‑embedding with vectoring sharing to eliminate redundant computations and DAG‑based batch pipelines with cost‑aware scheduling to minimize the inference time. Implemented as a PostgreSQL extension with LibTorch, MorphingDB outperforms AI‑native DBMSs (EvaDB, Madlib, GaussML) and AutoML platforms (AutoGluon, AutoKeras, AutoSklearn) across nine public datasets, encompassing series, NLP, and image tasks. Our evaluation demonstrates a robust balance among accuracy, resource consumption, and time cost in model selection and significant gains in throughput and resource efficiency.
\end{abstract}


\received{April 2025}
\received[revised]{July 2025}
\received[accepted]{August 2025}

\maketitle

\section{Introduction}


AI‑native DBMSs integrate deep learning inference into database engines to support advanced cognitive tasks such as face detection~\cite{DBLP:conf/ecai/0045GHCL23}, object recognition~\cite{DBLP:conf/cvpr/YueCGLGL24}, and entity resolution~\cite{di2019interpreting}.
These systems extend traditional relational and vector databases, including PostgreSQL~\cite{postgres}, Vexless~\cite{su2024vexless}, VectorDB~\cite{xie2025_vector}, and AquaPipe~\cite{yu2025aquapipe}, to provide unified data and model processing capabilities.
Existing implementations typically follow a \textbf{model‑centric} paradigm, obligating engineers to oversee the entire model lifecycle from architecture design and parameter tuning to deployment and version control~\cite{zhoudatabase}. This approach demands deep machine learning expertise and manual configuration of complex model details,  resulting in a substantial increase in  \textbf{development overhead}.


In contrast, several modern automated machine learning (AutoML) frameworks~\cite{karmaker2021automl, autosklearn, autokeras, autogluon} adopt \textbf{a task-centric paradigm} without requiring deep expertise in data science or machine learning. 
By iteratively sampling, training, and evaluating a pool of candidate models, these systems automatically identify the optimal model for any dataset and task. Motivated by this philosophy, we embed \textbf{the task‑centric module} into relational databases, enabling data engineers to develop complex AI applications via familiar SQL interfaces without manual model store, model selection, and hyperparameter tuning.


\begin{table*}[ht]
\caption{Two Design Paradigms}\label{tab:paradigm}
\begin{tabular}{p{0.1\linewidth}p{0.45\linewidth}p{0.37\linewidth}}
\hline 
\multicolumn{1}{c}{\textbf{Features}}  & \multicolumn{1}{c}{\textbf{Model-Centric}}   & \multicolumn{1}{c}{\textbf{Task-Centric}} \\ 
\hline
Operators & Create Model, Drop Model, Predict Model, Update Model, Prune Model, ... & Select Model for Task, Register Task, Predict Task \\ 
\hline
Definition & USING engine = 'mlflow', model\_name = 'folder\_name', mlflow\_server\_url = 'http://0.0.0.0:5001/' & CREATE TASK sentiment\_classifier (INPUT=Text\_Blob, OUTPUT in 'POS, NEG, NEU', Type ='Classification') \\ 
\hline
Query Interface & \textbf{SELECT} U.gender, U.location, AVG(\textit{TextCNN} ForSentiAnalysisV\_2\_0(R.comment)) \textbf{FROM} user \textbf{AS} U, Product \textbf{AS} P, review \textbf{AS} R \textbf{WHERE} U.ID=R.uid \textbf{AND} P.id=R.pid \textbf{AND} \textit{UNet}ForImageRecognitionV\_1\_2(P.img)= "iPhone 16" \textbf{AND} \textit{MaskRCNN}ForImage- RecoginitionV\_0\_2(R.img)="iPhone 16" \textbf{GROUP BY} U.gender, U.location & \textbf{SELECT} U.gender, U.location, AVG( sentiment\_classifier(R.comment)) \textbf{FROM} user \textbf{AS} U, Product \textbf{AS} P, review \textbf{AS} R \textbf{WHERE} U.ID=R.uid \textbf{AND} P.id=R.pid \textbf{AND} ImageRecognition(P.img)="iPhone 16" \textbf{AND} ImageRecognition(R.img)="iPhone 16" \textbf{GROUP BY} U.gender, U.location \\
\hline
\end{tabular}
\end{table*}

Table~\ref{tab:paradigm} provides a comparative analysis of the model-centric and task-centric paradigms, highlighting differences in operator sets, task definitions, and query interfaces.
In the \textbf{model‑centric} paradigm, executing the same analytical task across multiple datasets requires developers to explicitly \textbf{manage model storage and selection}. Each model must be registered with specific metadata, such as model name, version, input/output schema, and invocation interface.
Data developers are responsible for identifying suitable models for different data modalities, performing parameter tuning, and integrating these models into the query logic. 
For example, one might deploy a \textit{U‑Net}~\cite{ronneberger2015u} model to segment iPhone 16 product images, whereas \textit{Mask R‑CNN}~\cite{he2017mask} is employed to detect and classify user comments in iPhone 16 reviews, which heavily rely on domain expertise. 


In contrast, the \textbf{task-centric} paradigm abstracts away low-level model details through high-level task declarations such as \textit{ImageRecognition} or \textit{SentimentClassification}. Instead of requiring developers to explicitly specify which model to use, the system automatically resolves the task to the most appropriate model based on data characteristics, system context, and available resources. This eliminates the need for manual model selection and fine-tuning, significantly reducing the barrier to entry for non-expert users. Task-centric systems streamline AI integration workflows and scale more efficiently across diverse use cases.

While the task-centric paradigm offers significant abstraction and usability benefits, directly adopting it within relational databases introduces \textbf{some challenges}:

\begin{itemize}
    \item Model Storage: Relational databases such as PostgreSQL are designed for structured data and lack native support for unstructured model components such as neural network architectures. While extensions like \textit{pgvector} allow storage of flat vectors, they omit essential tensor shape metadata, making them incompatible with PyTorch tensors. Alternatively, API-based remote models offer convenient access to powerful closed-source LLMs without local deployment, but introduce challenges including external dependencies, unpredictable latency, and rate limitations.

    \item Model Selection: Traditional AutoML pipelines determine the most suitable model for a given task and dataset through \textbf{iterative training and evaluation of multiple candidates}. While effective, this process is \textbf{computationally expensive} and poorly suited for online inference scenarios that demand low latency and real-time responsiveness.
    \item Query Efficiency: When inference operators are interleaved with relational operators (e.g., joins, filters, aggregations), conventional query engines often fail to \textbf{optimize on these heterogeneous operations}. They lack awareness of \textbf{model execution costs}, \textbf{hardware-specific constraints} (e.g., GPU availability and batching), and \textbf{data dependencies} introduced by model invocation. These limitations hinder the system’s ability to achieve high-throughput inference at scale—especially in task-centric workloads where model inference is repeatedly applied over large datasets.
\end{itemize}




To address the above challenges, we build the system {\morphDB} to integrate the task-centric paradigm in DBMS and bridge the performance gap between DBMS and deep learning frameworks.

Firstly,  we design specialized model schemas and data types tailored for storing and managing deep learning models within relational databases.
We introduce the \textbf{Mvec} representation, which efficiently encodes multi-dimensional tensors with explicit shape information to enable lossless conversion and in-database manipulation with PyTorch tensors.
In addition to BLOB-based storage, {\morphDB} implements a \textbf{decoupled architecture model storage}, which separates model structure (e.g., network layers and operations) from its trained weights. This design facilitates partial loading and fine-grained inference execution, enhancing scalability and resource efficiency in model management.

Secondly, we define the model selection problem as a transfer learning task, employing a two-phase framework consisting of an offline embedding phase and an online model selection phase. 
The core idea is to build a transferability-related subspace, mapping the models and tasks into this subspace. Within this subspace, the distance between model vectors and task vectors signifies the level of transferability.
During the offline phase, we extract transfer patterns from performance matrices of candidate models across diverse historical tasks.
In the online phase, we develop a projection mechanism that aligns target tasks into the same subspace through feature-aware mapping.
This two-phase framework enables real-time model selection via efficient vector multiplication without extensive fine-tuning, saving resources and time overhead.

Finally, {\morphDB} introduces two core optimization techniques that enhance tensor manipulation and inference efficiency. (1) It employs an innovative approach that combines pre-embedding with SIMD to share vectorized data, thereby eliminating redundant embedding computations. (2) To maximize device utilization and scalability, {\morphDB} implements a pipeline-based batch inference strategy using a Directed Acyclic Graph (DAG) of operators. By leveraging adaptive CPU-GPU device placement and applying advanced pipeline techniques to optimize execution order, {\morphDB} streamlines complex inference workflows and minimizes bottlenecks.

We conduct a comprehensive comparative analysis against several prominent systems, EvaDB~\cite{evadb}, Madlib~\cite{madlib}, and GaussML~\cite{guoliang2024gaussml} on nine public datasets, encompassing series, NLP, and image tasks.
We also utilize the CIFAR dataset to evaluate the multi-dimensional performance of model selection in multiple popular AutoML frameworks such as AutoGluon~\cite{autogluon}, AutoKeras~\cite{autokeras}, and AutoSklearn~\cite{autosklearn}.
Our results indicate {\morphDB} consistently outperforms other approaches in both performance and usability, offering an intuitive interface that simplifies AI task execution. Compared to the AutoML framework, it has a better balance between accuracy, resource efficiency, and training time in model selection.

The paper is organized as follows: Section~\ref{sec:overview} presents the system overview. Sections~\ref{sec:model_manage}–\ref{sec:optimizations} cover model storage, selection, and optimization. Section~\ref{sec:evaluation} reports experiments, followed by related work in Section~\ref{sec:related_works} and conclusion in Section~\ref{sec:conclusion}.
\section{System Overview}
\label{sec:overview}

\subsection{Task-Centric Paradigm}
\label{subsec:task_centric}

{\morphDB} introduces a {\em task-centric} paradigm to simplify the incorporation of deep neural models within database-driven analytical dataflows.
Rather than requiring developers to explicitly construct and manage models, which is typical in model-centric systems, {\morphDB} enables users to define high-level inference tasks (e.g., face detection or sentiment analysis), allowing the system to handle model selection, configuration, and deployment automatically. 
This shift from model-centric control to task-centric specification not only reduces the barrier to entry for non-expert users, but also enables declarative, resource-efficient deployment of AI functionality directly within the DBMS.

Let \( T \) denote the set of tasks that users aim to accomplish with their data, and let \( M \) represent the set of models available in our repository. 
The task-centric paradigm is formalized as a function \( f: T \rightarrow M \) that selects the most appropriate model from \( M \) for each task in \( T \) based on performance and accuracy metrics.

By allowing users to specify what task to perform rather than \textit{how to perform it}, {\morphDB} aligns with the declarative nature of traditional DBMSs, where users define desired outcomes rather than detailed execution procedures. This abstraction significantly reduces development complexity and improves system adaptability, enabling {\morphDB} to more easily evolve with changing data workloads and analytical needs.

\subsection{{\morphDB} Overview}
\label{subsec:workflow}

\begin{figure}[t]
\centering
\includegraphics[width=0.48\textwidth]
{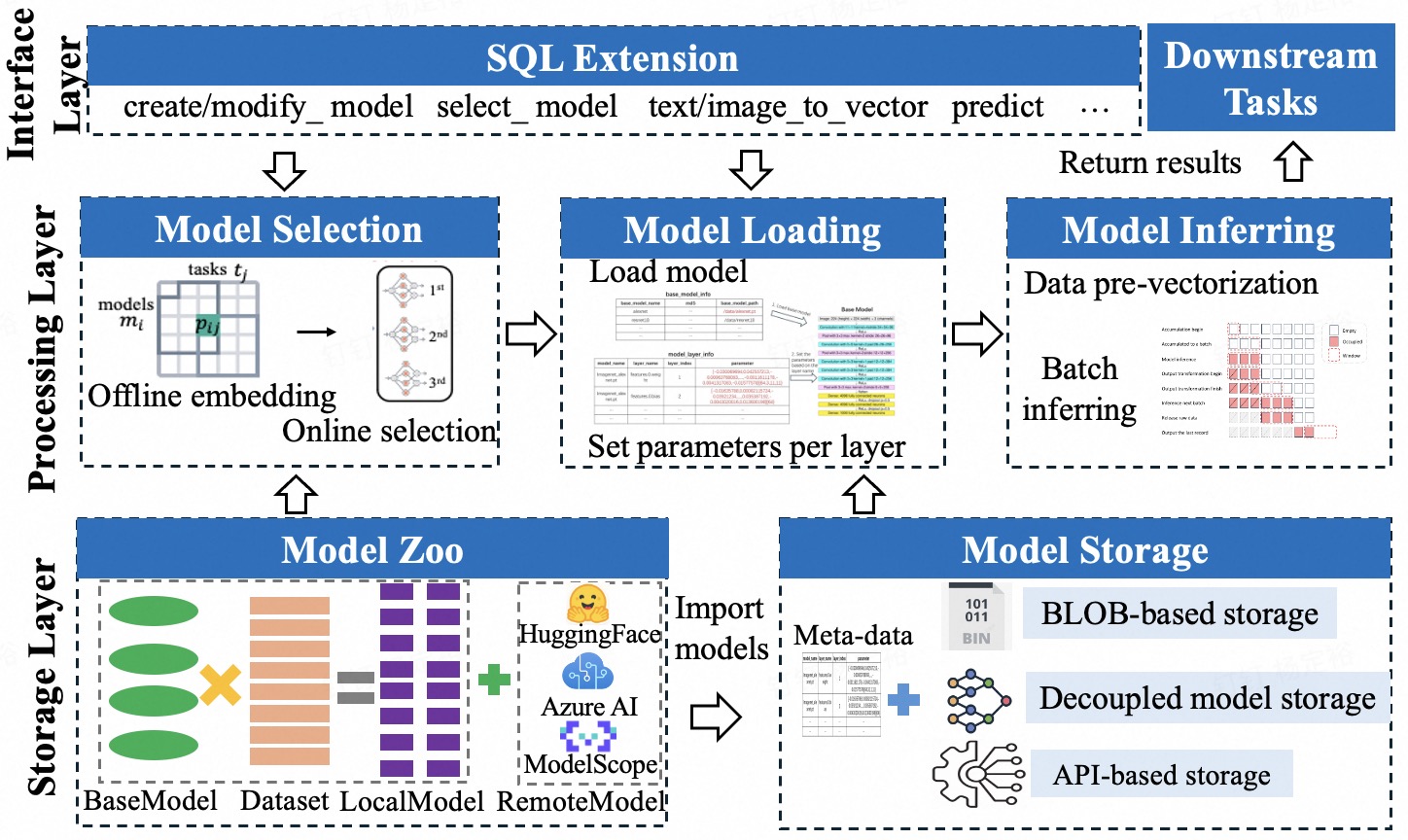}
 \caption{Overview of {\morphDB}}
 \label{fig:sys}
\end{figure}

The whole system is designed with a three-layer architecture to ensure efficient model management and inference execution. The storage layer handles the persistence of models and their associated metadata, enabling prompt retrieval and structured organization. The processing layer is responsible for orchestrating model selection, loading, and batch-based execution, optimizing computational efficiency and resource utilization. The interface layer provides an SQL-based interface for task inference, allowing users to integrate machine learning capabilities into database workflows.
A detailed overview of {\morphDB}’s primary workflow is illustrated in Figure~\ref{fig:sys}.

Within the storage layer, {\morphDB} maintains a unified \textbf{model zoo} encompassing both locally stored pre-trained models and externally hosted models, along with diverse datasets. Users can easily import models—either local or remote—into the system without requiring additional configuration. {\morphDB} automatically records model metadata in a centralized repository table.
To accommodate varying storage requirements, {\morphDB} supports multiple model formats: (1) \textbf{BLOB-based storage}, where both model architecture and parameters are stored as a single serialized object, (2) \textbf{Decoupled storage}, which separates model structure from parameters to enable selective loading, fine-grained updates, and better runtime flexibility.
(3) \textbf{API-based model integration}, where external endpoints can be registered as logical operators. This allows remote models to be seamlessly embedded into SQL queries alongside local models.


In the processing layer, {\morphDB} initiates the model selection process to identify the most appropriate pre-trained model from the model zoo. This process comprises offline task-model embedding to capture task-model relationships, followed by online model selection to dynamically choose the best model based on task requirements and performance metrics.
Once a model is selected, {\morphDB} loads it by retrieving the model metadata and progressively configuring layer-wise weight and bias parameters. Following this, our system executes the model inference process and generates results for downstream tasks.
Notably, throughout the designed pipeline, data vectorization for inference and batch inference optimizations are implemented to enhance inference efficiency.

The interface layer provides an intuitive and structured approach for users to define machine learning tasks through an extended SQL syntax.
To facilitate custom ML task definitions, this layer allows users to precisely specify input/output formats, expected neural model types, and performance constraints. By integrating these capabilities directly into SQL, {\morphDB} eliminates the need for low-level implementation details, making model deployment and inference more accessible.

{\morphDB} is implemented in C++ and integrated into PostgreSQL as an extension. 
The system incorporates LibTorch~\cite{libtorch}, the C++ version of PyTorch, to perform inference tasks. 
While a standalone Python library may simplify development, it hinders tight integration with the DBMS query engine, making it difficult to support operator-level scheduling, cost-based optimization, and SQL-native model invocation. Furthermore, Python-based solutions introduce runtime overhead, fragile dependencies, and complex deployments, limiting robustness and scalability in enterprise and cloud-native environments.


\noindent\textbf{Discussion.}
By abstracting inference as query operators and supporting heterogeneous backends (e.g., GPUs, remote APIs, decoupled models), {\morphDB} enables efficient, composable, and context-aware inference over both structured and unstructured data. 
This design naturally generalizes to LLM-based agentic systems: complex agents can be modularly decomposed into subtasks (e.g., retrieval, reasoning), mapped to LLMs of different capacities, and executed as directed acyclic graphs (DAGs). 
To optimize such workflows, {\morphDB}’s cost model can be extended for agent-aware planning, selecting models based on latency–accuracy–resource trade-offs. 
These capabilities position {\morphDB} as a foundation for AI-native agents that demand declarative task specification and fine-grained control over model execution across heterogeneous environments.

\begin{figure*}[t]
    \centering
    \subfloat[BLOB-based all-in-one model storage]
    {\includegraphics[width=0.38\textwidth]
    {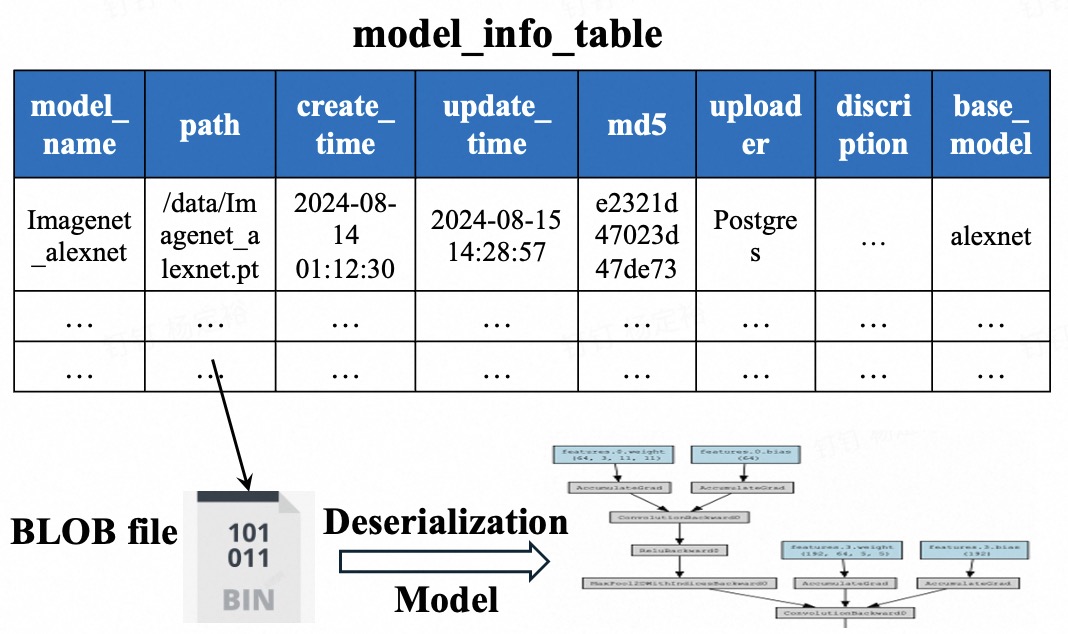}
    \label{fig:blob_model_store}}
    \hfill 
    \subfloat[Decoupled model storage]
    {\includegraphics[width=0.58\textwidth]
    {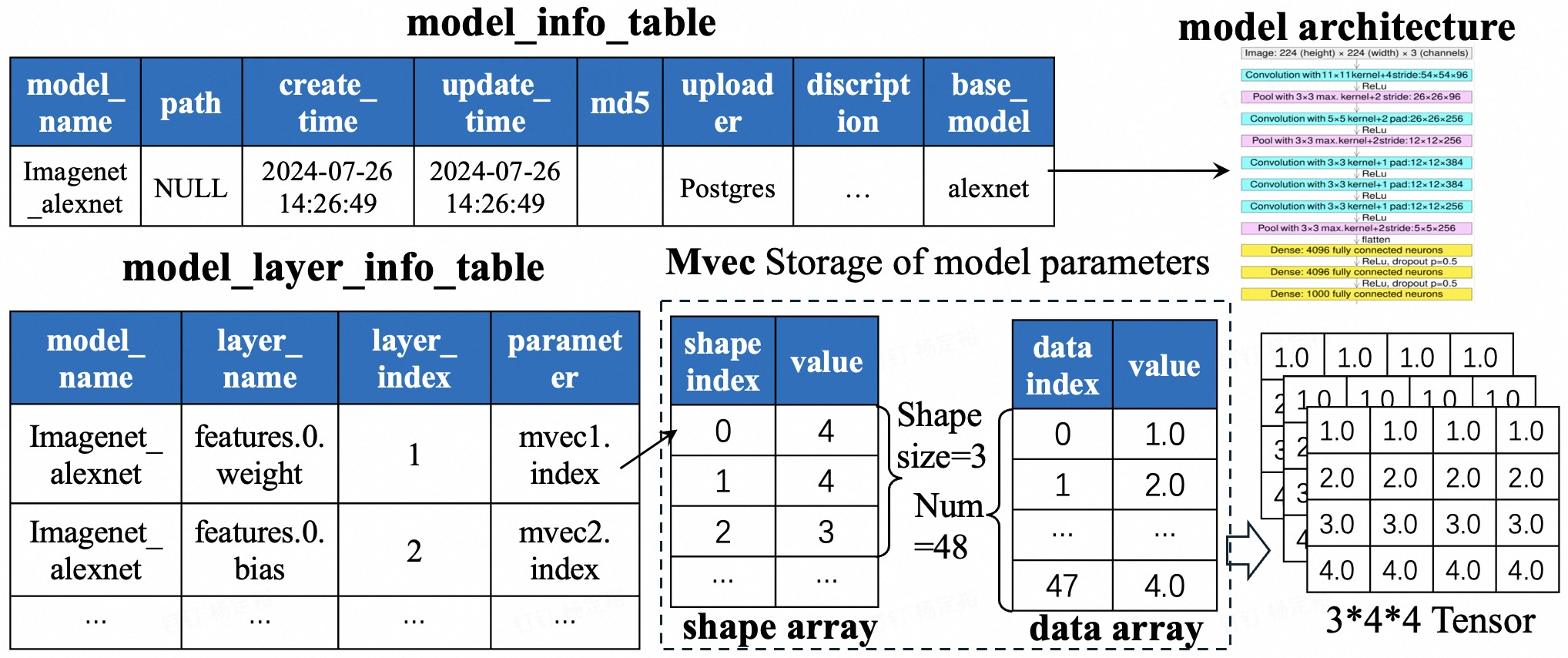}
    \label{fig:decoupled_model_store}}
    \caption{Model storage and management in {\morphDB}}
    \label{fig:store_model}
\end{figure*}

\section{Inner-DB Model Management}
\label{sec:model_manage}

In this section, we present how deep neural models are stored and managed in {\morphDB}. 
Specifically, we introduce a Mvec representation format to facilitate the storage for multi-dimensional
tensors.

\subsection{Model Storage}
\label{subsec:model_repo_manage}

We store the trained neural network models in a model repository for subsequent AI tasks and perform model inference computations by creating and loading models. 
A trained neural network model usually needs to save information such as network architecture, weight parameters, bias parameters, optimizer state, and other hyperparameters. 
{\morphDB} supports local model storage through two mechanisms: the binary large object (BLOB) file storage and the decoupled storage. In addition, it provides API-based integration to support remote or externally hosted models, including proprietary or closed-source services~\cite{openai_gpt4}.

\noindent\textbf{BLOB-based model storage.} 
This strategy serializes both the neural network architecture and its parameters into a single binary large object.
Model metadata, such as identifier, version, and storage path, is maintained in a {\em model\_info\_table} shown in Figure~\ref{fig:blob_model_store}. 
Although this design leverages the file system for straightforward implementation, it incurs substantial deserialization overhead during model loading, undermining low‑latency inference. Its monolithic format also restricts fine‑grained component loading—hindering pipeline parallelism, scalability, and the reuse of shared architectural elements—and disallows partial updates to model parameters. Consequently, {\morphDB} reserves BLOB-based storage for static models with minimal update or sharing requirements.

\noindent\textbf{Decoupled storage of model architecture and parameters.} 
To enable data reuse, partial model updates, and partial model loading, {\morphDB} proposes decoupled model storage, which extracts the model network architecture as the base model, such as ResNet18, ResNet50, AlexNet, etc., from the model weight and bias parameters of different layers, and stores them separately. 
As illustrated in Figure~\ref{fig:decoupled_model_store}, we also use the {\em model\_info\_table} to record the model metadata, and the {\em base\_model} field to record the pointers to the base architecture. 
An extra {\em model\_layer\_info} table is used to record the information of each model layer, including layer name, layer index, and the weight and bias parameters. 
Decoupled storage is similar to ONNX-model storage~\cite{onnx, onnx_in_oracle}, which is an open format focused on model exchanges for various LLM models and different platforms.
While storing base model weights and fine-tuned diffs externally in formats such as ONNX or LoRA-style deltas is feasible, it presents practical trade-offs. External storage introduces additional runtime dependencies and latency, and can complicate integration with in-DBMS planning and scheduling, potentially undermining optimization opportunities and system reliability.

\noindent\textbf{API-based model storage.}
{\morphDB} abstracts external model endpoints as logical operators within its task-centric execution framework. Each API-based model is registered with metadata such as endpoint URL, input/output schema, expected latency, authentication credentials, and usage quotas. 
During query planning, these operators are treated equivalently to local models through HTTP/REST or gRPC interfaces with automatic request construction, payload serialization, and result deserialization.
To ensure robustness and performance, {\morphDB} includes adaptive retry logic, timeout control, and response caching. 

\noindent\textbf{Discussion.}
{\morphDB} abstracts over heterogeneous model integration strategies—ranging from in-database to API-based deployment—within a unified task execution framework. 
This design allows users to flexibly select the most appropriate integration method based on deployment constraints, workload characteristics, and performance objectives. Some advantages and limitations are summarized as follows:
\begin{itemize}
    \item BLOB-based storage: This method stores both the model architecture and parameters as a single binary object, offering simplicity and fast loading for small to medium-sized models. It is ideal for static or single-tenant scenarios where models rarely change and update granularity is not required,  but lacks reuse flexibility and incurs redundancy across variants.
    \item Decoupled storage: {\morphDB} separates architecture metadata from weight tensors, supporting selective loading, fine-grained updates, and partial inference execution. This reduces redundancy across fine-tunes (e.g., ResNet-50 variants) and is well-suited for multi-tenant, memory-constrained, or high-throughput settings.
    \item API-based integration: This mode enables easy access to powerful, closed-source LLMs without requiring local deployment or fine-tuning, making it ideal for lightweight tasks, exploratory queries, or rapid prototyping. However, it introduces external dependencies, variable latency, and limited optimization within the query planner.
\end{itemize}

\subsection{Mvec Tensor Representation}
\label{subsec:model_tensor}

While formats like ONNX are widely used for model representation and exchange, they are not optimized for fine-grained, database-native storage and access patterns.
To enable in‑database storage of multi‑dimensional model parameters and direct interoperability with PyTorch, we introduce a new tensor format-\textbf{Mvec representation}. 
Unlike PostgreSQL’s \textit{pgvector} extension, which only accommodates one‑dimensional vectors by storing raw values and vector length without shape metadata, Mvec preserves both the tensor data and its full dimensionality, enabling accurate, lossless conversions between database‑resident tensors and PyTorch tensors.
Each tensor in Mvec is represented by two contiguous arrays (Figure~\ref{fig:decoupled_model_store}): 
\begin{itemize}
    \item A \textbf{shape array}, which records the size of each tensor dimension (e.g., channels, height, width);
    \item A \textbf{data array}, that stores the tensor's elements as a flattened, one-dimensional sequence in row-major order, meaning that elements in the same row are stored contiguously before moving to the next row.
\end{itemize}

When ingesting a PyTorch tensor, {\morphDB} serializes its dimensions into the shape array and its values into the data array. 
Conversely, to reconstruct the original tensor, the system reads the shape array to determine dimensionality, computes stride information to map one‑dimensional indices back to multi‑dimensional coordinates, and reshapes the data array accordingly. 
MVec is designed specifically to support shape-aware binary tensor storage tightly integrated with the DBMS, enabling efficient SQL-level filtering, slicing, and partial loading of tensor data—operations that ONNX does not natively support.
By mirroring PyTorch’s contiguous memory layout, Mvec ensures efficient storage, rapid I/O, and seamless tensor reconstruction within the DBMS.

\subsection{Model Zoo} 
\label{subsec:model_zoo}

To support heterogeneous task requirements across diverse data modalities, {\morphDB} incorporates a modular and extensible \textbf{Model Zoo}, which serves as the backbone for automated, task-driven model selection. This repository is populated with a rich collection of pretrained models sourced from established model hubs such as the PARC benchmark~\cite{parc}, Hugging Face~\cite{hugging_face}, and Ollama~\cite{ollama}, enabling out-of-the-box support for a wide range of AI tasks without requiring users to collect datasets, fine-tune models, or procure GPU resources.
For instance, ResNet variants~\cite{he2020resnet} such as ResNet-18/34 are optimized for edge scenarios like camera-based defect detection, while ResNet-50/101 serve higher-accuracy tasks like facial recognition and object tracking. Similarly, the YOLO family~\cite{ge2021yolox} offers a spectrum from lightweight versions (e.g., YOLOv5-nano) for mobile inference to larger variants (e.g., YOLOv5-x, YOLOv8) for industrial-grade detection. These variants differ in architecture, fine-tuning data, and deployment profiles, underscoring the need for systems like {\morphDB} to manage diverse model variants efficiently and contextually.

The Model Zoo is organized by task type and data modality, supporting image classification, time series analysis, and natural language processing, each with distinct input characteristics, data distributions, and semantic structures. For image tasks, the Zoo includes tiered convolutional architectures from the PARC benchmark~\cite{parc}, such as ResNet-50, ResNet-18, and AlexNet, each pretrained on canonical datasets like ImageNet or CIFAR-10 to ensure broad generalization. For time series applications, we integrate lightweight MLP-based models trained across datasets with varying temporal granularity~\cite{ekambaram2023tsmixer}, supporting regression and classification objectives. In the NLP domain, we include transformer-based encoders such as ALBERT~\cite{2020ALBERT} and RoBERTa~\cite{RoBERTa}, fine-tuned for tasks such as domain-specific sentence classification and semantic matching.

Unlike conventional model-centric frameworks that require explicit user configuration of architectures and hyperparameters, {\morphDB} adopts a task-centric interface: model selection is guided by high-level task semantics and dataset properties. The Model Zoo thus functions not only as a storage layer for local pretrained models or remote API-based models, but also as a substrate for automated matching during inference planning, enabling efficient, domain-adaptive deployment without user intervention.

Currently, the Model Zoo contains 198 models across the three primary modalities (image, time series, text), and is designed to support future extensions through plug-in architectures and continual ingestion of models. This design ensures that {\morphDB} remains adaptable to evolving workloads, providing a robust foundation for unified, multi-task AI computation within relational databases.

\section{Model Selection}
\label{sec:model_select}

To address the high cost of iterative training and evaluation across multiple models and datasets, {\morphDB} employs a two-phase model selection strategy. First, an offline embedding phase constructs a transferability-aware subspace that captures relationships between tasks and models. Then, during the online phase, the system performs real-time model selection through lightweight vector operations without requiring feature extraction or fine-tuning, substantially reducing resource consumption and latency.

\subsection{Problem definition.}
\label{subsubsec:select_offline}

Adopting the most appropriate model for different tasks can essentially be seen as a transfer learning problem. 
Although the matching score can be estimated using the correlation between the label and the forward features, the time complexity of this approach grows linearly with the number of models. 
This paper delves into the intricate relationship between tasks and models, introducing a novel methodology to reduce time overhead. 
The core idea is to build a transferability-related subspace, mapping the models and tasks into the subspace.
Within this subspace, the distance between model vectors and task vectors signifies the level of transferability.

Formally speaking, given a target task $t^*$, our goal is to select a suitable model $m^*$ from the 
massive model zoo $\mathcal{M}=\{ m_i\}_{i=1}^M$ so that the model $m^*$ can achieve best performance on $t^*$. 
To this end, we compute a transfer score $\text{Trans}(m_i, t^*)$ between each model $m_i$ and the target task $t^*$, as shown in Equation~\ref{eq:trans_score}.
We solve this problem by offline model/task embeddings and online model selection. 
\begin{equation}
m^* = \arg\max_{m_i \in \mathcal{M}} \text{Trans}(m_i, t^*).
    \label{eq:trans_score}
\end{equation}

\begin{figure}[th]
    \centering
    \includegraphics[width=0.9\linewidth]{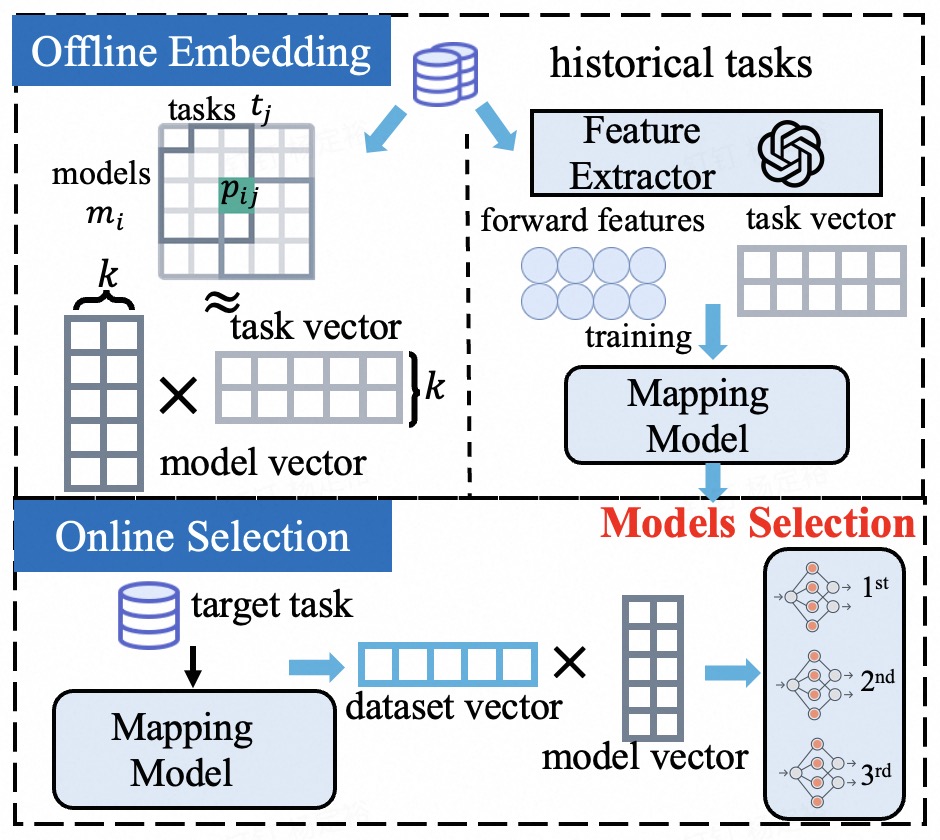}
    \vspace{-0.3cm} 
    \caption{Model selection based on task embeddings}
    \label{fig:model_selection}
\end{figure}

\subsection{Offline: model and task embeddings.}
\label{subsubsec:select_offline}
To construct a subspace related to transfer capabilities for mapping models and tasks, we initially collect the real transfer performance of historical tasks $\mathcal{T}=\{ t_j\}_{j=1}^N$ on the models in the model zoo $\mathcal{M}$ during the offline phase.
The historical transfer matrix is denoted as $V \in \mathbb{R}^{M\times N}$, where $N$ is the number of historical tasks. 
Each element $v_{ij} \in V$ represents the performance of model $i$ on task $j$. 
Subsequently, we extract latent transfer patterns representing the performance of models across various historical tasks from the $V$.

The goal of matrix factorization is to decompose a matrix into multiple low-dimensional matrices, which reside in the $k$-dimensional subspace. The embeddings in the latent subspace represent the underlying features of the data, capturing significant patterns within it. In our scenario, the $k$-dimensional subspace can be regarded as latent factors influencing model-to-task transferability. 
Specifically, we decompose this matrix with the objective function in Equation~\ref{eq:mf}, where $W \in \mathbb{R}^{M\times k}$ and $H \in \mathbb{R}^{N\times k}$ respectively represent the embeddings of models and historical tasks. The inner product of model embedding $\boldsymbol{m_i} \in W $ and task embedding $\boldsymbol{t_j} \in H$
reflects the matching degree between them.
\begin{equation}
    \begin{aligned}
       & \min_{W,H}  \ \| V - WH^\top \|_F^2 \\
        & \text{subject to} \ W, H \geq 0
    \end{aligned}
    \label{eq:mf}
\end{equation}
where \( \| \cdot \|_F \) denotes the Frobenius norm.

\subsection{Online: model selection.}
\label{subsubsec:select_online}

In the online phase, our goal is to select the most suitable model for a target task $t^*$. 
Since the embeddings of historical tasks lie in the learned $k$-dimensional latent space, selecting a model requires the embedding $\boldsymbol{t^*}$ of the new task.
However, directly computing this embedding is non-trivial as $t^*$ was not involved during training.

To address this, we leverage the powerful generalization capabilities of Large Vision Models (LVMs), such as CLIP, which are pre-trained on a wide range of vision and text tasks. 
Specifically, we collect forward features of historical tasks using the LVM and pair them with their learned embeddings $H \in \mathbb{R}^{N \times k}$. Using this data, we train a regression model $R$ to learn the mapping from the LVM feature space to the latent task embedding space.
We adopt the CLIP model as the forward feature extractor and a random forest as the regressor for training, as shown in Equation~\ref{eq:clip}.
The reason we adopt CLIP is that CLIP supports both image and text modalities, equipping our approach with the ability to perform model selection across different modalities.
This approach is grounded in the assumption that tasks exhibiting similar LVM features tend to share similar transferability characteristics, enabling the regressor to generalize to unseen tasks.
\begin{equation}
    \boldsymbol{t_j} = R_{random\_forest}(CLIP.encode\_image(t_j))
    \label{eq:clip}
\end{equation}

The embedding of target task $\boldsymbol{t^*}$ can be deduced by passing its LVM forward features to the regressor $R$. 
Finally, the transfer scores $\text{Trans}(m_i, t^*)$ of the models in the repository on the target task can be obtained by multiplying the learned model embeddings with the task embedding.
As shown in Equation~\ref{eq:rank}, we can directly rank the estimated transfer scores to select the most suitable model $m^*$. 
This model selection method reduces the need for feature extraction and extensive fine-tuning on the target task for each model in the model repository, significantly saving resources and time overhead involved in model selection.
\begin{equation}
    m^* = \arg\max_i ( \text{Trans}(m_i, t^*) )  =  \arg\max_i ( \boldsymbol{m_i} * \boldsymbol{t^*} \mid \boldsymbol{m_i} \in H )  
    \label{eq:rank}
\end{equation}

\section{Inner-DB inference Optimizations}
\label{sec:optimizations}

To bridge the performance gap between DBMS and ML frameworks, 
{\morphDB} incorporates two core optimization techniques designed to enhance tensor manipulation and inference efficiency: optimized vectorized sharing, and pipelined batch inference.

\subsection{In-Database Vectorized Sharing}
\label{subsec:vectorization}

In traditional inference pipelines, each query involving images, text, or other modalities triggers an on-the-fly embedding process, followed by model-specific inference. As a result, even when the same dataset is analyzed repeatedly, the system redundantly computes embeddings for each query, leading to unnecessary overhead and increased latency.

We find that the feature extraction stage—responsible for converting raw data into vectorized embeddings—is decoupled from the subsequent inference stage.
Once data is embedded, the resulting vectors are independent of the particular inference model used, which allows them to be effectively reused across diverse downstream tasks.
Therefore, we propose a pre-embedding and sharing mechanism to mitigate the inefficiencies caused by repeatedly embedding data for diverse user tasks and inference operations.

\begin{figure}
    \centering
    \includegraphics[width=1\linewidth]{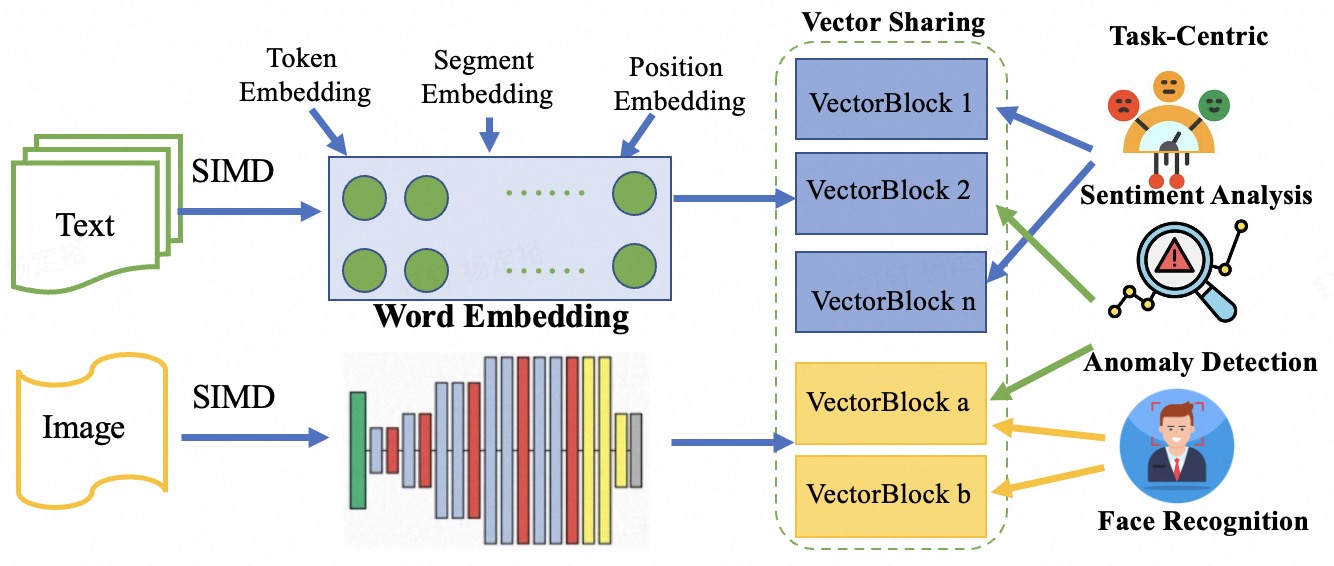}
    \caption{In-Database Vectorization Sharing}
    \label{fig:vector_sharing}
\end{figure}

The process begins with a dedicated feature extraction pipeline that converts raw data into high-dimensional feature vectors optimized for database storage, as illustrated in Figure~\ref{fig:vector_sharing}. 
For text data, we employ pre-trained language models such as ALBERT~\cite{2020ALBERT} or BGE embedding~\cite{BGE} to generate semantic embeddings. For image data, widely adopted convolutional neural network architectures, including ResNet~\cite{krizhevsky2012imagenet}, VGG~\cite{vgg2015}, and Inception~\cite{szegedy2016rethinking}, are used to extract compact vector representations. These embeddings are stored in specialized database tables and organized into vector blocks that support various AI inference tasks.

To further enhance performance, we integrate SIMD (Single Instruction, Multiple Data) instructions into the vectorization process. Rather than processing image pixels sequentially, where each pixel is individually scaled and transformed, our approach processes groups of pixels concurrently. For example, during image normalization, multiple pixel values are loaded simultaneously into SIMD registers, which perform parallel arithmetic operations to apply the necessary transformations across the entire group in a single operation. An analogous strategy is applied to text data, where simultaneous processing of multiple token embeddings accelerates the generation of semantic representations.
This batch processing strategy minimizes computational overhead for individual queries and enables the precomputation of embeddings that can be efficiently reused across multiple inference tasks, thereby optimizing overall system throughput and scalability.

\subsection{Batch Inference in Pipeline}
\label{subsec:batch}
Previous research~\cite{postgres, xing2024database} has seen the development of AI databases within PostgreSQL that cater to various machine-learning tasks. 
However, these databases often face challenges in inference efficiency, particularly when dealing with intricate queries and extensive datasets–a hurdle also encountered by ML researchers in their early stages. 
While some efforts have been made to modify the PostgreSQL core, these modifications can pose difficulties during kernel upgrades, limiting the incorporation of community features. 
To overcome the efficiency shortcomings without the need for core modifications to PostgreSQL, {\morphDB} introduces a batch inference pipeline strategy for AI databases. 
This strategy aims to maximize device utilization and scalability to enhance overall performance.

\begin{figure}
    \centering
    \includegraphics[width=1\linewidth]{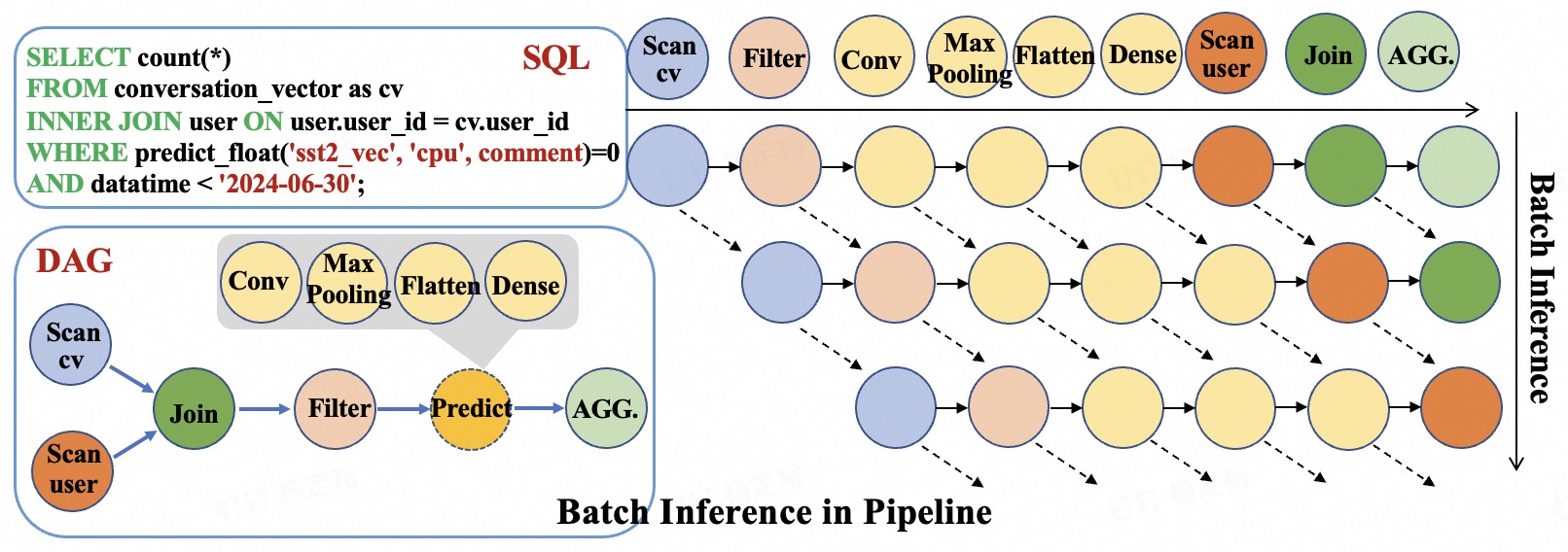}
    \caption{The pipeline processing of batch inference}
    \label{fig:batchinf_1}
\end{figure}

Figure \ref{fig:batchinf_1} illustrates an example of our batch inference pipeline framework, which contains multiple relational algebra operators (i.e., $JOIN$, $FILTER$) and model inference function $Predict$. 
{\morphDB} first parses the SQL statement into a series of operators. 
These operators are structured as a Directed Acyclic Graph (DAG), where each node represents an individual operation (e.g., $SCAN$, $JOIN$, $FILTER$), and the edges denote dependencies between them. 
If the query has AI inference functions, our inference model is recognized as a specialized node in the DAG and then fetches the model structure, weights, and hyperparameters stored in the database. For example, an Alexnet model has to assemble the parameters of $Conv$, $MaxPooling$, $Flatten$, and $Dense$ Layers into an executable inference module. 
The assembly of inference functions allows the system to adapt to different models and task requirements, enhancing flexibility and scalability in the inference pipeline.

\noindent\textbf{CPU+GPU Co-processing:}
As shown in Figure \ref{fig:batchinf_1}, one complex SQL query might integrate relational algebra operators and AI model inference functions, within a database system optimized for AI tasks. 
This requires overcoming the inherent differences in computational characteristics between relational algebra operators (e.g., $JOIN$, $FILTER$, $AGGREGATE$) and AI inference functions (e.g., $Predict$, $Sentiment Classifier$). Relational operators typically rely on CPU-bound execution due to their control flow-heavy nature and the need for efficient handling of random memory access and branching logic. These operators are generally optimized on multi-core CPUs, where techniques such as multi-threading can accelerate operators like filtering, joining, and aggregating large datasets. 
Traditional AI inference tasks also perform reasonably well on CPUs, especially for models with lower computational operators. However, when handling more complex deep learning networks, such as convolutional neural networks (CNNs) or transformers, the latency introduced by CPU-bound execution becomes prohibitively high.
Our framework supports leveraging GPUs to accelerate these compute-intensive inference operators like large-scale matrix multiplications, convolutions, and attention mechanisms, which require high levels of parallelism and memory bandwidth.

The execution of heterogeneous operators introduces overhead from data movement between CPU and GPU memory spaces, especially when combined with network latency incurred by remote model access. 
These memory and network transfers can significantly impact end-to-end performance, adding latency that may offset the computational gains of GPU acceleration. Therefore, optimizing execution plans requires careful orchestration to balance GPU throughput with memory synchronization and remote communication costs, ensuring that the benefits of hardware acceleration and model integration justify the overhead incurred.

We propose a cost model to determine the optimal execution device (CPU or GPU) for both traditional relational operators and AI inference operators within an SQL query pipeline. 
The total cost of an operator $op$ is estimated as:
\begin{equation}
    C_{op} = ExecTime_{op} + TransCost_{op} 
    \label{equation:cpu_gpu_model}
\end{equation}
where $ExecTime_{op}$ denotes the operator's local or remote execution time, and $TransCost_{op}$ accounts for data transfer time—either between CPU and GPU memory spaces or across the network when invoking remote models. Notably, for external, closed-source models where $ExecTime_{op}$ is not directly observable, we approximate $C_{op}$ using the end-to-end response latency.


To instantiate the unified cost model, we provide separate formulations for GPU and CPU executions, considering the architectural and memory hierarchy differences between the two devices.

For GPU-based execution, the total cost $C_{GPU}$ accounts for both the computation time $ExecTime_{GPU}$ and the data transfer overhead $TransCost_{GPU}$ incurred during model invocation:
\begin{equation}
ExecTime_{GPU} = \frac{ModelFLOPS}{FLOPS} \times nrows
\end{equation}
\begin{equation}
TransCost_{GPU} = \frac{ModelSize}{MemBW} + \frac{ModelSize}{GPUBW} + \text{Latency}
\end{equation}
where $ModelFLOPS$ denotes the total floating-point operations required by the model, $FLOPS$ reflects the peak computational throughput of the GPU, 
$nrows$ is the number of input samples processed or inferred in the batch, 
$ModelSize$ refers to the size of the model in bytes,
$MemBW$ represents the bandwidth of main memory,
$GPUBW$ measures the transfer speed from main memory to GPU memory, 
and $Latency$ encompasses I/O or network latency.

In contrast, CPU execution avoids explicit memory hierarchy transitions, leading to a simplified cost expression:
\begin{equation}
ExecTime_{CPU} =  \frac{ModelFLOPS}{FLOPS} \times nrows
\end{equation}
\begin{equation}
TransCost_{CPU} = \frac{ModelSize}{MemBW}
\end{equation}

The key distinction is that CPU execution avoids memory–GPU transfer, making it more efficient for small tasks, while GPUs offer higher throughput for compute-intensive workloads.

\textbf{Device Selection Strategy:}
To determine the optimal execution device, we adopt a cost-based decision framework derived from the unified performance model:
\begin{equation}
\text{Device} = \begin{cases}
\text{GPU} & \text{if } C_{GPU} < C_{CPU} \\
\text{CPU} & \text{otherwise}
\end{cases}
\end{equation}

This criterion ensures that the execution platform is selected based on a quantitative assessment of end-to-end cost, incorporating both computation time and data transfer time. By leveraging this analytical model, the system can dynamically allocate operators to the most suitable device, adapting to variations in model size, data volume, and system bandwidth.

\noindent\textbf{Pipeline Processing:}
To further improve throughput during inference tasks, {\morphDB} employs advanced pipeline techniques to streamline the execution of complex inference workflows. One critical challenge is identifying dependency relations between various operators, ensuring that tasks are processed in an optimal order without introducing bottlenecks in the DAG. We propose a dependency analysis algorithm based on the DAG to optimize the execution flow. The details are depicted in Algorithm \ref{algo:dependency}.
The algorithm first constructs a dependency map of operators by analyzing the edges between nodes (lines 3-5) and classifying dependency relationships between operators (lines 6-12). 
It then schedules the operators for execution by performing a topological sort using a Depth First Search (DFS) approach (lines 13–15).
This method allows the system to prioritize operators with higher computational demands or critical operators, facilitating effective task scheduling and parallelization.

\begin{algorithm}[t]
\begin{algorithmic}[1]
\caption{The Algorithm of Pipeline Dependency Discovery}
\label{algo:dependency}
\State \textbf{Input:} DAG $G = (V, E)$ where $V$ is the set of operators, and $E$ represents dependencies
\State \textbf{Output:} Execution order $\sigma$
\ForAll{$v \in V$}
    \State $\mathcal{D}(v) \gets \{ u \in V \mid (u, v) \in E \}$
\EndFor

\ForAll{$(u, v) \in E$}
    \If{is\_data\_dependency$(u, v)$}
        \State Label $(u, v)$ as "Data Dependency"
    \Else
        \State Label $(u, v)$ as "Control Dependency"
    \EndIf
\EndFor
\ForAll{$v \in V$}
    \State $\sigma$ $\gets$ \Call{DFS}{$v$} 
\EndFor
\State{\Return $\sigma$} 

\end{algorithmic}
\end{algorithm}

\noindent\textbf{Batch Inferences:}
The inference function is reformulated as a window function to improve the throughput, with the parameter \textit{batch\_size}. 
The introduction of \textit{batch\_size} allows for flexibility in the inference process. When set to 1, the operator function is similar to a non-batch inference operator that processes one data point at a time.

To support the batch inferences in PostgreSQL, we modify the kernel's window function in three aspects: 
(1) \textbf{Window Data Aggregation}: Raw data from underlying PostgreSQL operators is duplicated into an intermediate state to extend its lifecycle. This ensures data is retained long enough for batch processing to improve efficiency during inferences.
(2) \textbf{Batch Inference Execution}: Once a batch is filled, the system triggers the inference process. The raw data is processed in parallel using a thread pool to convert into input tensors. These tensors are combined along the batch dimension, maximizing throughput and resource utilization by leveraging hardware optimizations in model inference.
(3) \textbf{Data Cleanup and Result Caching}: After completing inference, the results are transformed back into a PostgreSQL-compatible format, with tensors and other data types cached in an intermediate state for continued processing. Previous raw data is efficiently released in this phase, and parallelization ensures that this cleanup step sustains high throughput across batch operations.

Designing an optimal batch size for inferences is crucial to maximizing throughput and maintaining efficient resource utilization. Smaller batch sizes offer higher concurrency but lead to suboptimal use of computational resources, while larger batch sizes improve throughput but reduce concurrency due to higher memory. To address this challenge, {\morphDB} designs a cost model that considers several factors to automatically select an optimal batch size. The model will balance the trade-off between throughput (T) and latency (L) while accounting for resource utilization (R), such as memory, computational power (e.g., CPU or GPU), and concurrency (C). Given a batch size $B$, the cost $C(B)$ can be calculated as:
\begin{equation}
    C(B) = \arg\max_B(F(L(B), R(B), C(B), T(B)))
    \label{equation:batch_size_model}
\end{equation}

To delve deeper into the implications of different batch size configurations, we conduct experimental studies to analyze the impact of varying batch sizes, as detailed in Section \ref{subsec:exp_ablation}. 
Our experimental findings reveal that batch size settings ranging from 8 to 32 typically yield optimal inference efficiency.

\section{Evaluation}
\label{sec:evaluation}

\subsection{Experiment Setup}
\label{subsec:exp_setup}

In this section, we detail the datasets, methodologies, and hardware configurations for evaluating the performance of our system. 
Moreover, the {\morphDB} codebase is publicly available at: \hyperlink{https://github.com/MorphingDB/MorphingDB}{https://github.com/MorphingDB/MorphingDB}.

\begin{table}[H]
  \caption{Nine datasets in three domains}
  \label{tab:datasets}
  \begin{tabular}{clcr}
    \toprule
    Type & Dataset & \#Dimension & Dataset Size \\
    \midrule
    \multirow{4}{*}{Series} & YearPredict\cite{year_prediction_msd_203} & 90 cols & 515,345 rows \\
    & Slice\cite{relative_location_of_ct_slices_on_axial_axis_206} & 384 cols & 53,500 rows \\
    & Swarm\cite{swarm_behaviour_524} & 2400 cols & 24,017 rows \\
    \midrule
    \multirow{3}{*}{Image} & ImageNet\cite{krizhevsky2012imagenet} & 3*224*224 & 14,197,122 images \\
    & CIFAR-10\cite{Krizhevsky09learningmultiple} & 3*224*224 & 60,000 images \\
    & Stanford Dogs\cite{Khosla2012NovelDF} & 3*224*224 & 20,580 images \\
    \midrule
    \multirow{3}{*}{NLP} & SST-2\cite{socher-etal-2013-recursive} & 4*128 & 70,042 sentences \\
    & IMDb\cite{maas-EtAl:2011:ACL-HLT2011} & 4*128 & 50,000 sentences \\
    & Financial\cite{Malo2014GoodDO} & 2*133 & 4,840 sentences \\
    \bottomrule
  \end{tabular}
\end{table}

\textbf{Datasets:} 
We employ nine widely-used open datasets to benchmark the performance of AI-powered databases in Table \ref{tab:datasets}. These datasets encompass three distinct types of inference tasks: Series tasks, Image tasks, and NLP tasks. 
The YearPredict dataset \cite{year_prediction_msd_203} includes audio features paired with song release years from 1922 to 2011. The Slice dataset \cite{relative_location_of_ct_slices_on_axial_axis_206} contains features from CT scans with a numeric class indicating slice position within the body. The Swarm dataset \cite{swarm_behaviour_524} captures high-dimensional features of animal swarm behavior, classified into binary motion categories. 
We also perform image inference tasks such as image classification on ImageNet\cite{krizhevsky2012imagenet}, CIFAR-10\cite{Krizhevsky09learningmultiple}, Stanford Dogs\cite{Khosla2012NovelDF} datasets.
SST-2\cite{socher-etal-2013-recursive}, IMDb\cite{maas-EtAl:2011:ACL-HLT2011}, Financial\cite{Malo2014GoodDO} are popular datasets for NLP inferences, i.e., binary sentiment classification tasks.

\textbf{Hardware configuration:} 
We utilize an Intel(R) Xeon(R) Gold 6240 CPU @ 2.60GHz with 16 threads and 32GB of memory as the CPU server to evaluate efficiency and performance using the CPU. Additionally, we assess the efficiency and performance using an NVIDIA RTX 3090 GPU with 24GB of memory, on an Intel(R) Xeon(R) Silver 4316 CPU @ 2.30GHz with 40 threads and 32GB of memory as the GPU server.

\textbf{Baselines:} 
We compare several open-source AI-native databases that integrate machine learning capabilities to support various algorithms. For example, functional integration system EvaDB~\cite{evadb}, kernel-level integration systems Madlib~\cite{madlib} and GaussML~\cite{guoliang2024gaussml} are considered for efficient model prediction. 
We also consider whether the platforms are open-source, reproducible, and structurally comparable to {\morphDB}. For instance, MindsDB is excluded since MindsDB and EvaDB are similar in that both serve as middleware frameworks with ML frameworks.
In addition, we evaluate our model selection strategy against widely adopted open-source AutoML platforms. AutoKeras~\cite{autokeras} leverages Neural Architecture Search to automatically identify optimal deep learning models and hyperparameters.
AutoSklearn~\cite{autosklearn} performs the selection and tuning of machine learning algorithms for a variety of tasks, while AutoGluon~\cite{autogluon} processes multiple data modalities with GPU acceleration to handle more complex tasks.

\textbf{Workloads:}
We define flexible, general-purpose templates (detailed in our Supplement File) within the database to support various AI inference workloads, including series classification and regression, NLP sentiment classification, and image classification tasks. 
For each dataset, we construct a representative set of tasks aligned with the dataset’s characteristics and systematically generate 100+ queries using predefined label configurations. 
For time-series regression tasks, the query workload is relatively lightweight, primarily comprising selection, projection, window functions, and prediction. In comparison, NLP sentiment analysis and image classification introduce higher computational complexity by incorporating additional operations such as aggregation, group-by, and inference operators, which increase processing overhead and intermediate state management. Furthermore, multi-modal inference tasks further extend this complexity by performing cross-modal data fusion through join operations across heterogeneous inputs (e.g., text and image), followed by window functions applied to the integrated feature space. 
We run each query at least three times and take the average to ensure the robustness and consistency of the evaluation. 
However, when we observe higher variance (e.g., $>$20\%)—often due to dynamic resource contention—we perform additional trials to ensure statistical stability and report representative averages.

\textbf{Structure of Experimental Trials:}
To comprehensively evaluate {\morphDB}'s performance across a broad range of scenarios, particularly to validate its robustness and efficiency under diverse task modalities, including series, NLP, and image inference workloads. We conduct multiple groups of experiments:
(1) \textbf{Overall performance evaluation} assesses the integrated behavior of {\morphDB} under nine datasets across three domains: series, NLP, and image, to validate the effectiveness and robustness in different scenarios. (2) \textbf{Component-level analysis} focuses on core modules such as model storage, model selection, and CPU-GPU Placement, requiring controlled trials to isolate the effect of each mechanism. (3) For \textbf{ablation studies}, we select one representative dataset per domain (e.g., Slice for time series, SST-2 for NLP, CIFAR-10 for image) to control the complexity and space constraints.

\subsection{Efficiency and performance Study}
\label{subsec:efficiency&performance}

In this section, we conduct a comprehensive efficiency and performance evaluation of {\morphDB} across three domains using nine benchmark datasets. The results are illustrated as follows.

\subsubsection{Series Tasks}
\label{subsec:series_tasks}

\begin{figure}[tp]
    \centering
    \captionsetup[subfigure]{skip=4pt}
    \begin{subfigure}{0.49\linewidth}
        \centering
        \includegraphics[width=\linewidth]{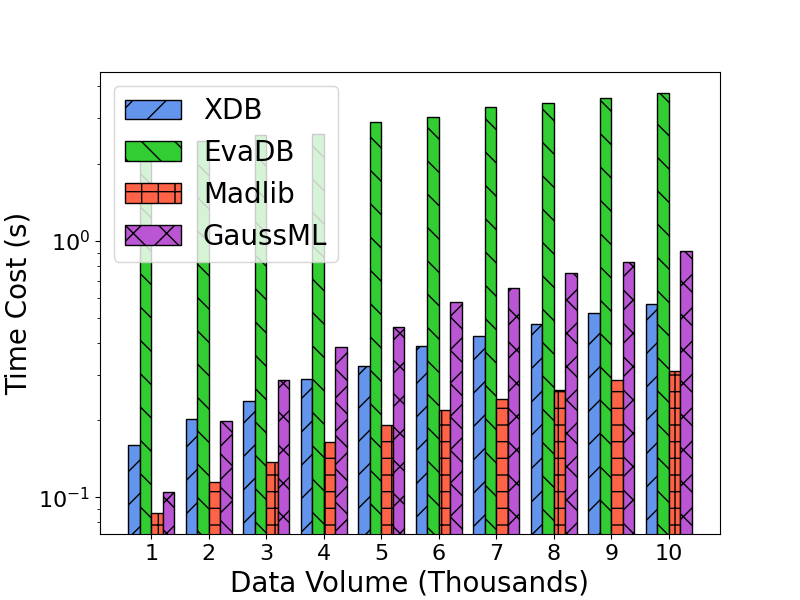}
        \caption{YearPredictMSD}
        \label{fig:series_yearpredict}
    \end{subfigure}
    \hfill
    \begin{subfigure}{0.49\linewidth}
        \centering
        \includegraphics[width=\linewidth]{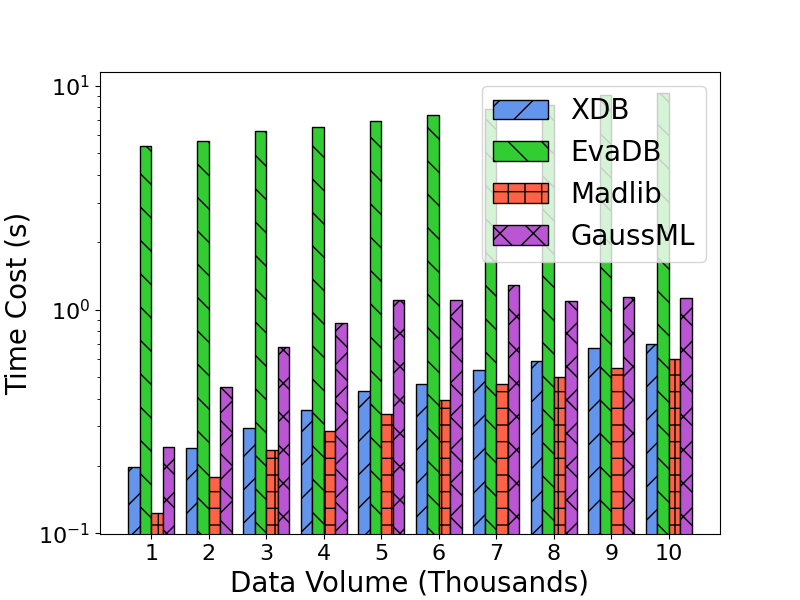}
        \caption{Slice}
        \label{fig:series_slice}
    \end{subfigure}
    \hfill
    \begin{subfigure}{0.49\linewidth}
        \centering
        \includegraphics[width=\linewidth]{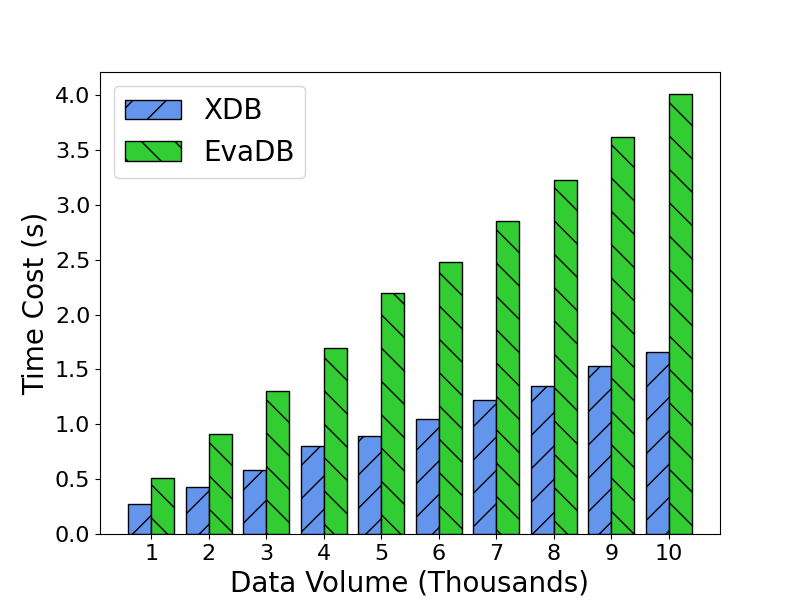}
        \caption{Swarm}
        \label{fig:series_swarm}
    \end{subfigure}
    \hfill
    \begin{subfigure}{0.49\linewidth}
        \centering
        \includegraphics[width=\linewidth]{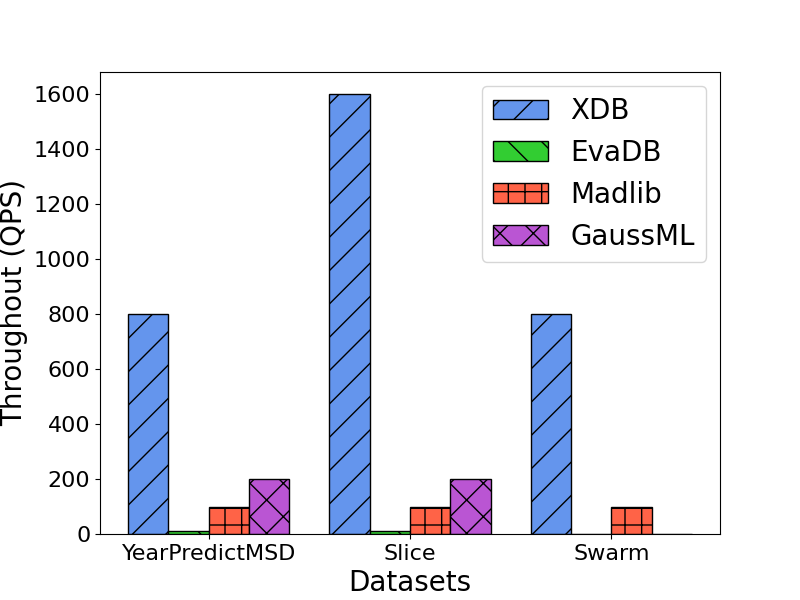}
        \caption{Throughput}
        \label{fig:series_throughput}
    \end{subfigure}
    \caption{Performance comparison on Series Tasks}
    \label{fig:series}
\end{figure}

We compare the performance with EvaDB, GaussML, and Madlib on series inference tasks across three datasets with different data volumes.
Figure~\ref{fig:series_yearpredict} and~\ref{fig:series_slice} illustrate the performance of \textbf{series classification tasks} on \textit{YearPredic} dataset (90 columns) and \textit{Slice} dataset (380 columns), respectively. 
As data volume increases, inference time rises across all systems in both cases.
EvaDB demonstrates the highest inference time for both datasets, indicating less efficient handling of classification tasks on low-dimensional data.
Madlib and {\morphDB}, however, show close efficiency, suggesting both systems are well-suited to moderate-dimensional data, while GaussML lags slightly due to its less optimized query execution and data access methods.
We also study the performance of \textbf{regression tasks} for the \textbf{high-dimensional} Swarm dataset (2400 columns) shown in Figure~\ref{fig:series_swarm}. Notably, both EvaDB and GaussML encountered issues with PostgreSQL’s column limit on the Swarm dataset, resulting in execution errors. 

Figure~\ref{fig:series_throughput} analyzes the throughput of {\morphDB} and other systems, and we can find that {\morphDB} achieves at least a \textbf{4x higher} throughput compared to other systems due to its implementation of batch inference within a pipelined architecture. 
{\morphDB} achieves the highest throughput, even though its per-query latency is slightly higher than Madlib. This is primarily because {\morphDB} employs optimized batch inference pipelines (e.g., DAG-based operator scheduling and vector sharing), which allow it to amortize overhead across multiple concurrent queries and maximize hardware utilization. 
GaussML, despite its longer average latency, GaussML's throughput improvements primarily benefit from internal optimizations such as native operator fusion and are driven by hardware-level acceleration for specific workloads.
By processing batches of data in parallel, {\morphDB} minimizes the idle time between stages, maximizing hardware utilization and reducing the overall latency per query.

\subsubsection{NLP Tasks}
\label{subsec:nlp_tasks}

\begin{figure}
    \centering
    \captionsetup[subfigure]{skip=4pt}
    \begin{subfigure}{0.49\linewidth}
        \centering
        \includegraphics[width=\linewidth]{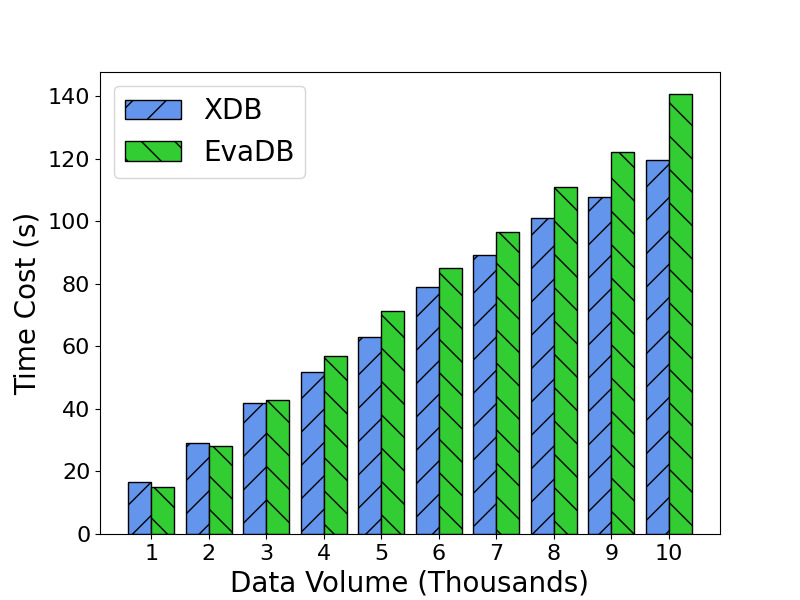}
        \caption{Finance}
        \label{fig:finance}
    \end{subfigure}
    \hfill
    \begin{subfigure}{0.49\linewidth}
        \centering
        \includegraphics[width=\linewidth]{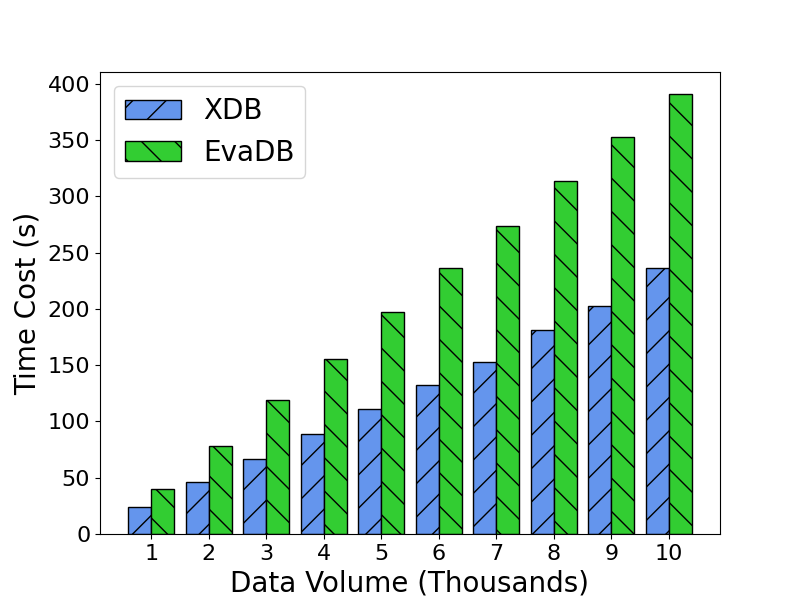}
        \caption{SST-2}
        \label{fig:sst_2}
    \end{subfigure}
    \hfill
    \begin{subfigure}{0.49\linewidth}
        \centering
        \includegraphics[width=\linewidth]{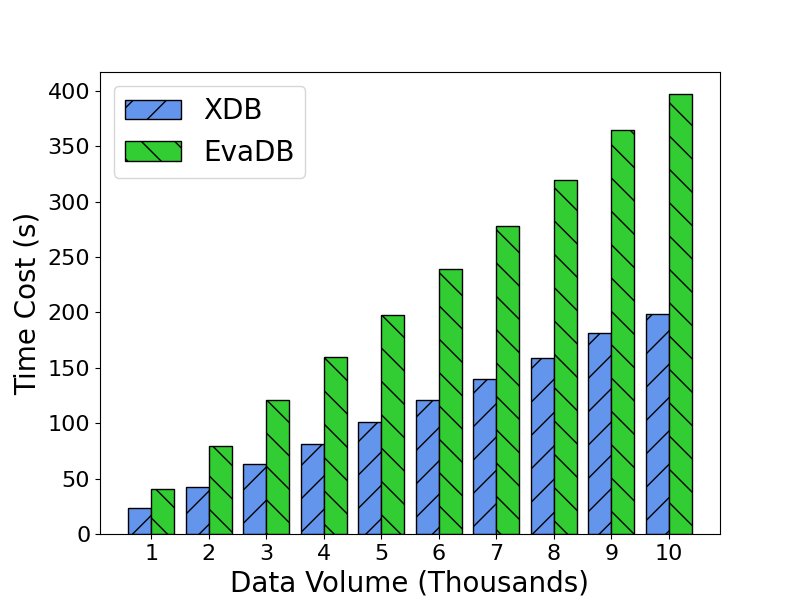}
        \caption{IMDB}
        \label{fig:imdb}
    \end{subfigure}
    \hfill
    \begin{subfigure}{0.49\linewidth}
        \centering
        \includegraphics[width=\linewidth]{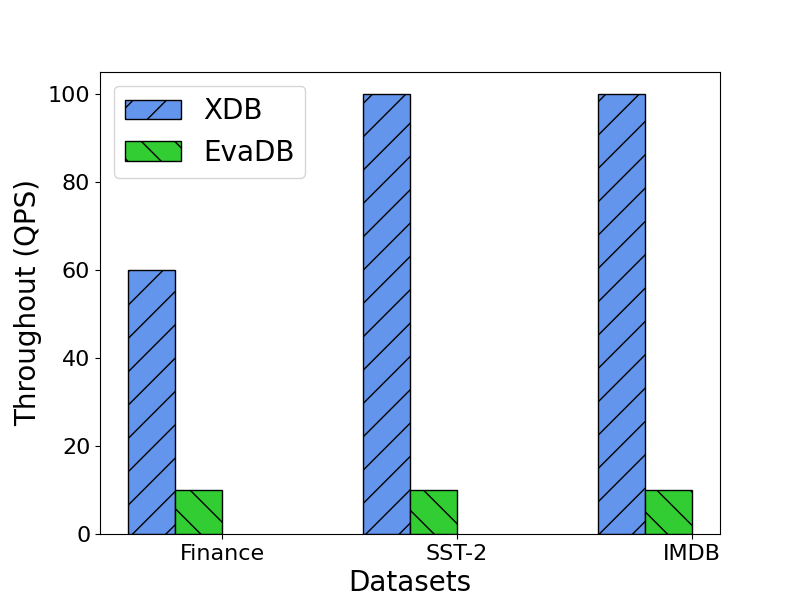}
        \caption{Throughput}
        \label{fig:nlp_throughput}
    \end{subfigure}
    \caption{Performance comparison on NLP Tasks}
    \label{flg:nlp}
\end{figure}

We evaluate the inference efficiency of natural language processing tasks, specifically focusing on \textbf{sentiment classification} using the \textit{ALBERT} model. Due to the lack of support for this model in GaussML and Madlib, EvaDB is selected as our baseline for comparison. The experiments are conducted with a batch size of 16 to ensure consistent evaluation conditions for both systems. 
As shown in Figure \ref{flg:nlp}, {\morphDB} demonstrates a significant improvement in inference efficiency across the three datasets, achieving a relative enhancement ranging from 10\% to 50\% compared with EvaDB, largely because of the vector sharing and batch pipeline optimization.

Notably, in the \textit{Finance} dataset in Figure~\ref{fig:finance}, the performance advantage of {\morphDB} is less pronounced. This can be attributed to the Finance dataset consists primarily of shorter sentences, reducing computational complexity and minimizing the processing time difference between {\morphDB} and EvaDB. However, {\morphDB} still has a higher throughput than EvaDB shown in Figure~\ref{fig:nlp_throughput}.
These findings indicate {\morphDB}’s capability to handle NLP tasks with particularly strong performance on longer, more complex text inputs.

\subsubsection{Image Tasks}
\label{subsec:image_tasks}

\begin{figure}
    \centering
    \captionsetup[subfigure]{skip=4pt}
    \begin{subfigure}{0.49\linewidth}
        \centering
        \includegraphics[width=\linewidth]{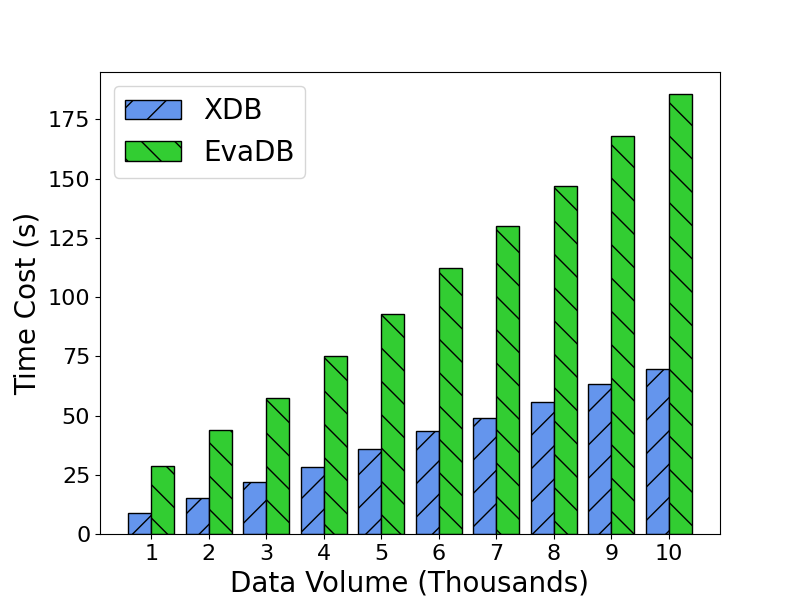}
        \caption{Stanford Dogs}
        \label{fig:stanford_dogs_a}
    \end{subfigure}
    \hfill
    \begin{subfigure}{0.49\linewidth}
        \centering
        \includegraphics[width=\linewidth]{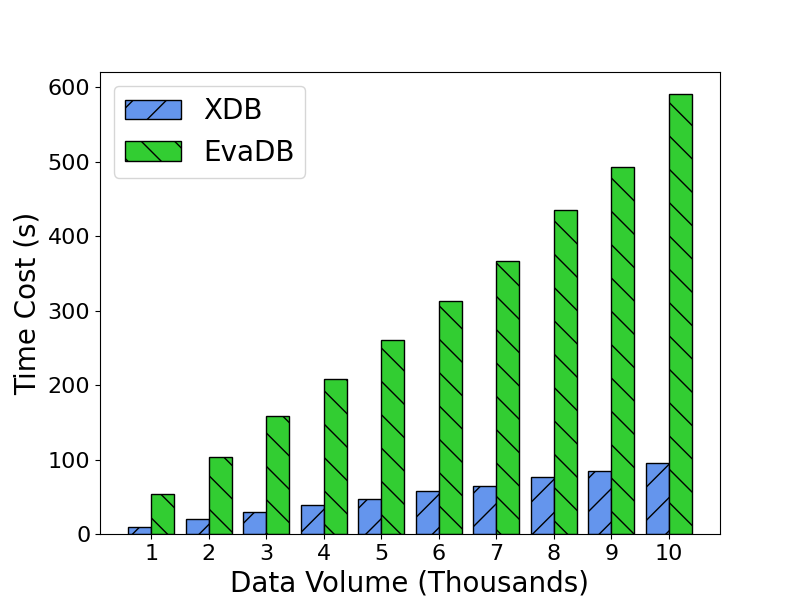}
        \caption{ImageNet}
        \label{fig:stanford_dogs_b}
    \end{subfigure}
    \hfill
    \begin{subfigure}{0.49\linewidth}
        \centering
        \includegraphics[width=\linewidth]{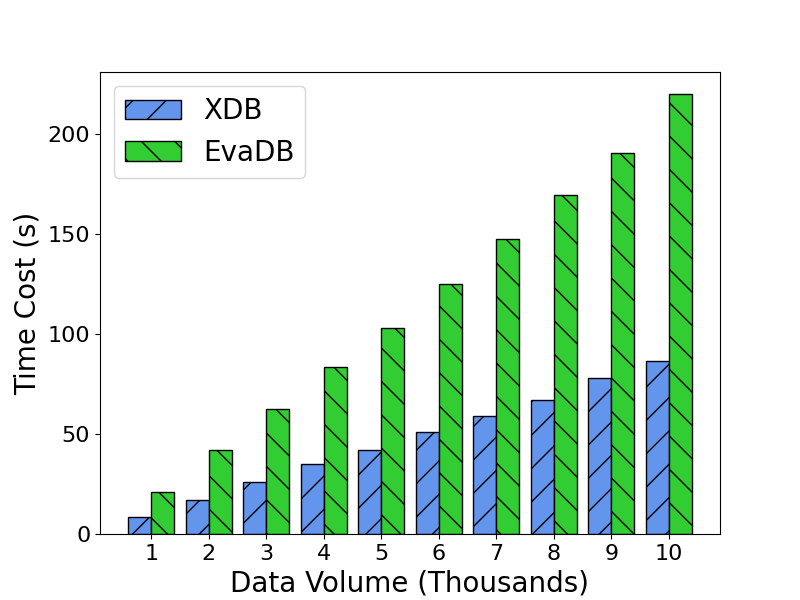}
        \caption{CIFAR-10}
        \label{fig:stanford_dogs_c}
    \end{subfigure}
    \hfill
    \begin{subfigure}{0.49\linewidth}
        \centering
        \includegraphics[width=\linewidth]{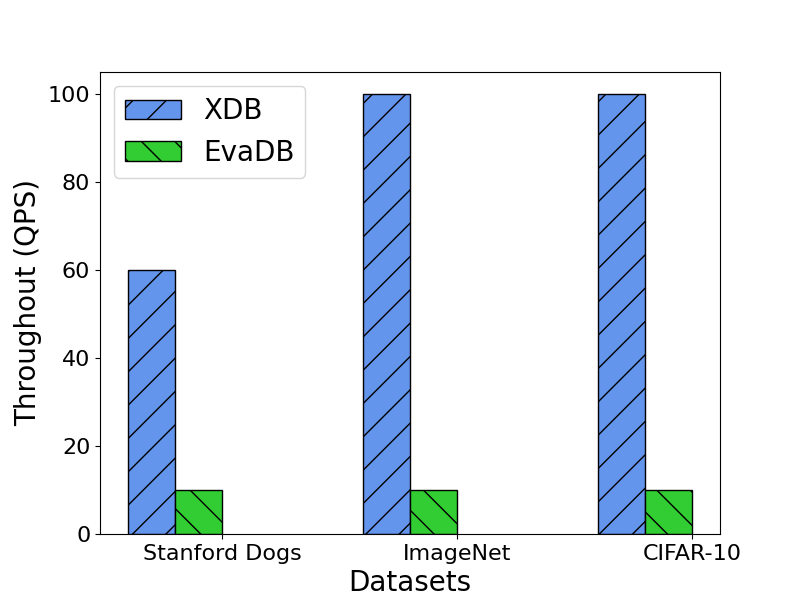}
        \caption{Throughput}
        \label{fig:image_throughput}
    \end{subfigure}
    \caption{The performance comparison on Image datasets}
    \label{fig:dataset_comparison}
\end{figure}

We further evaluate the \textbf{image classification} efficiency of {\morphDB} using the AlexNet, ResNet-18, and GoogleNet models across three benchmark datasets: Stanford Dogs, ImageNet, and CIFAR-10. Since GaussML and Madlib lack support for image classification tasks, we only compare {\morphDB} and EvaDB. 
As shown in Figure~\ref{fig:dataset_comparison}, in comparison with EvaDB, {\morphDB} achieves a substantial reduction in inference time, with an average decrease exceeding 70\% across the three datasets.
This performance advantage is primarily due to {\morphDB}’s storage of pre-embedding feature vectors directly within the database, contrasting with EvaDB’s reliance on raw binary image data stored outside a database system. By leveraging PostgreSQL’s optimized data management, {\morphDB} significantly reduces retrieval and access latency, which contributes to faster data handling and processing.

For instance, on the \textit{Stanford Dogs} dataset, {\morphDB} achieves a substantial efficiency gain over EvaDB. This improvement is attributed to {\morphDB}'s vectorized storage format, which not only enables efficient organization and retrieval of image embeddings but also supports optimized inference on high-dimensional data. In contrast, EvaDB relies on image data storage, which incurs additional retrieval and decoding overhead, which heavily impacts performance on larger image datasets. 

\subsection{Effect of Model Storage}
\label{subsec:model_management_study}
We conduct a comparative analysis of the model storage strategies: In-Database models, and API-based/external models. Specifically, we assess their storage usage, model loading overhead, and inference time associated with each approach.

\begin{figure}
    \centering
    \captionsetup[subfigure]{skip=4pt}
    \begin{subfigure}{0.32\linewidth}
        \centering
        \includegraphics[width=\linewidth]{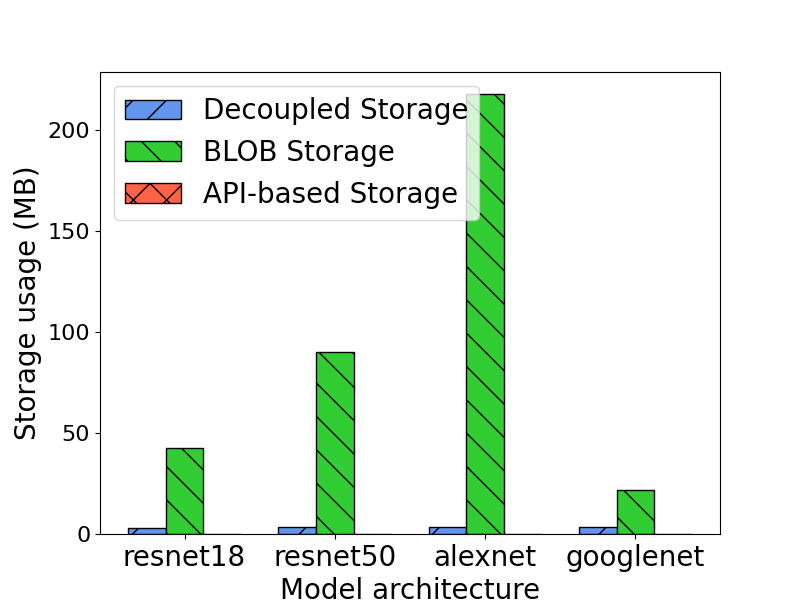}
        \caption{Storage Usage}
        \label{fig:disk_usage}
    \end{subfigure}
    \begin{subfigure}{0.32\linewidth}
        \centering
        \includegraphics[width=\linewidth]{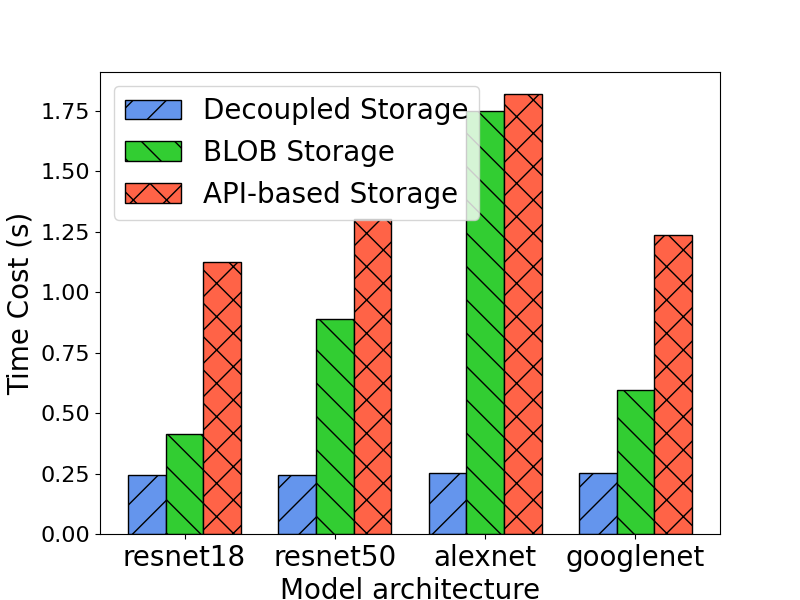}
        \caption{Loading time}
        \label{fig:load_time}
    \end{subfigure}
    \begin{subfigure}{0.32\linewidth}
        \centering
        \includegraphics[width=\linewidth]{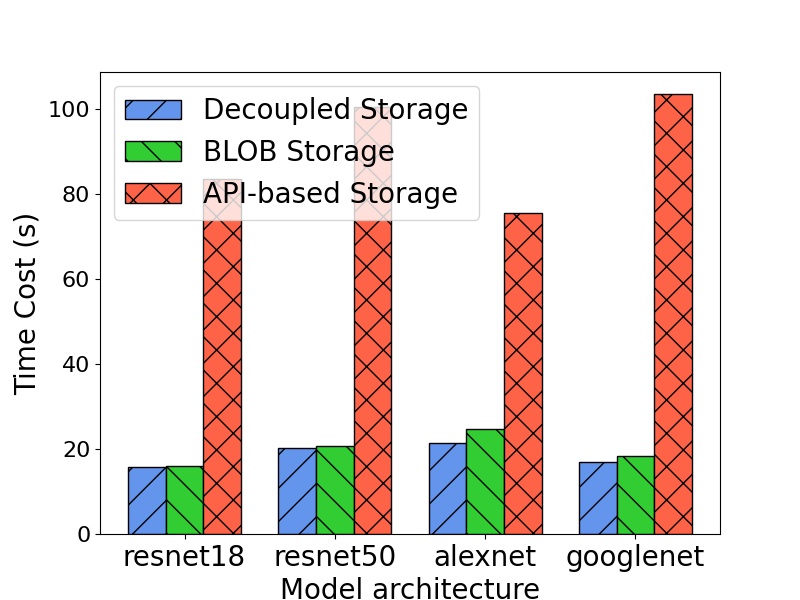}
        \caption{Inference time}
        \label{fig:infer_time}
    \end{subfigure}    
    \caption{Comparison of model storage strategies}
    \label{fig:model_storage_comparison}
\end{figure}

\textbf{Storage Usage}.
Figure~\ref{fig:disk_usage} illustrates the model storage usage
for different storage types.
We find that BLOB-based storage results in the highest disk usage, as it stores the entire model—including both architecture and parameters. Decoupled storage, by contrast, only records the updated layers while reusing the base model, thus significantly reducing redundancy and storage cost. In comparison, API-based methods have negligible storage overhead, as the model is deployed externally and only minimal metadata (e.g., URL, version) needs to be stored locally.

\textbf{Model loading time}.
Figure~\ref{fig:load_time} compares the model loading time without inference across different storage strategies. BLOB-based models have a higher loading time, as the entire model binary must be deserialized and reconstructed at once. 
Decoupled storage mitigates this bottleneck by separating the architecture and parameter files, allowing partial loading and even parallel access to different model components. This significantly reduces initialization latency, especially in dynamic or resource-constrained environments. For the API-based method, it mainly involves lightweight service preparation, such as metadata retrieval and connectivity checks.

\textbf{Model inference time}.
Figure~\ref{fig:infer_time} presents the inference performance when processing 10,000 records on a GPU. API-based models incur the highest latency, primarily due to the overhead of HTTP-based communication with external services. 
Both BLOB-based and decoupled in-database models operate locally, resulting in significantly lower inference times. 

\subsection{Effect of Model Selection}
\label{subsec:automl_model_selection}

\begin{figure}
    \centering
    \captionsetup[subfigure]{skip=4pt}
    \begin{subfigure}{0.32\linewidth}
        \centering
        \includegraphics[width=\linewidth]{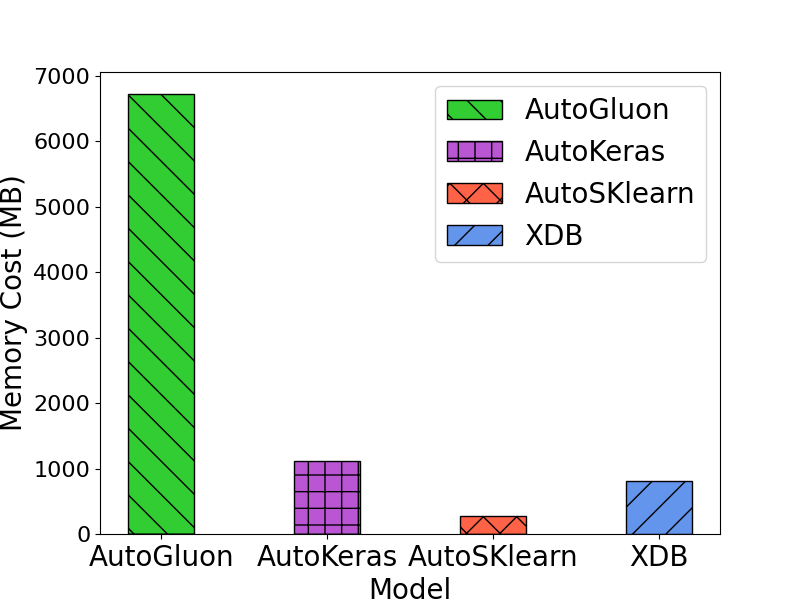}
        \caption{Memory Cost}
        \label{fig:Automl_mem_cost}
    \end{subfigure}
    \begin{subfigure}{0.32\linewidth}
        \centering
        \includegraphics[width=\linewidth]{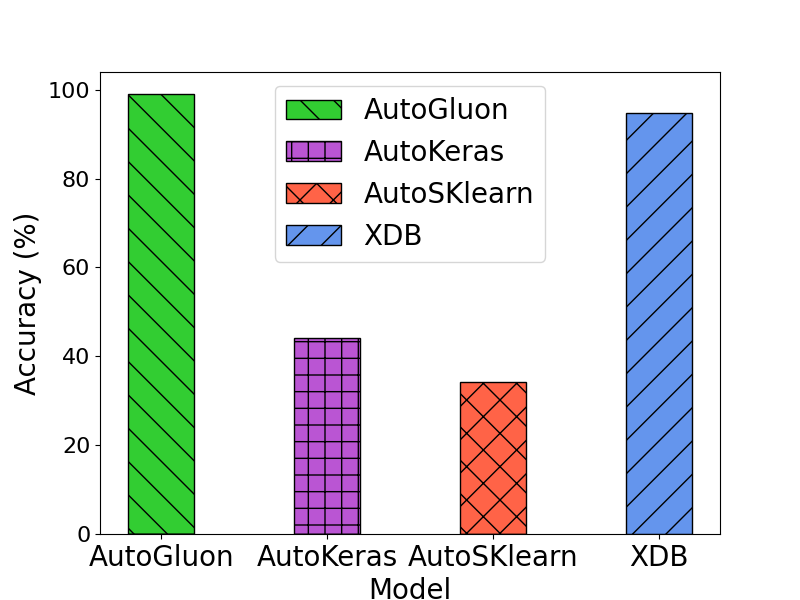}
        \caption{Accuracy}
        \label{fig:Automl_accuracy}
    \end{subfigure}    
    \begin{subfigure}{0.32\linewidth}
        \centering
        \includegraphics[width=\linewidth]{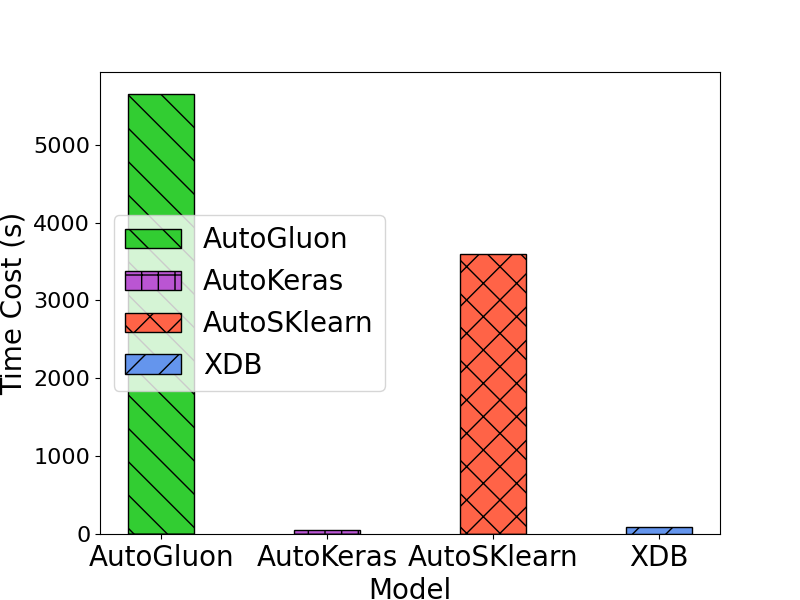}
        \caption{Time Cost}
        \label{fig:Automl_time_cost}
    \end{subfigure}
    \caption{Comparison of AutoML platforms}
    \label{fig:Automl_evaluate}
\end{figure}

We compare our model selection algorithm of {\morphDB} against AutoGluon, AutoKeras, and AutoSklearn in terms of resource consumption, processing time, and accuracy. Figure~\ref{fig:Automl_evaluate} presents a comparative evaluation based on the CIFAR-10 dataset.
\textit{AutoGluon} achieves the highest accuracy among the evaluated platforms, demonstrating its capability to identify optimal model architectures and hyperparameters through the two-stage selection. However, this accuracy is attained at the cost of substantial computational resources—both in terms of memory usage and extended training durations. Consequently, this method is highly effective in resource-rich environments where accuracy is paramount.
\textit{AutoSklearn} offers a resource-efficient alternative, consuming significantly less memory compared to its counterparts. Despite this advantage, its accuracy is comparatively lower, which can restrict its utility in applications requiring high precision. Moreover, the training time remains relatively high, despite its lower memory usage.
\textit{AutoKeras} provides a balanced approach by delivering rapid training times alongside moderate memory usage. This makes it particularly attractive for applications where speed is critical. Nonetheless, its accuracy does not match that of AutoGluon or {\morphDB}, suggesting that AutoKeras is more suitable for preliminary analyses or tasks where immediate results are prioritized over achieving the highest possible accuracy.
{\morphDB} demonstrates a robust balance among accuracy, resource consumption, and time cost. By leveraging transfer learning, {\morphDB} incorporates pre-trained models that are subsequently fine-tuned for specific tasks. This strategy significantly reduces processing time and memory usage compared to models trained from scratch, while still achieving high accuracy and making it particularly well-suited for environments where both performance and resource constraints are critical.

\begin{figure}
    \centering
    \captionsetup[subfigure]{skip=4pt}
    \begin{subfigure}{0.32\linewidth}
        \centering
        \includegraphics[width=\linewidth]{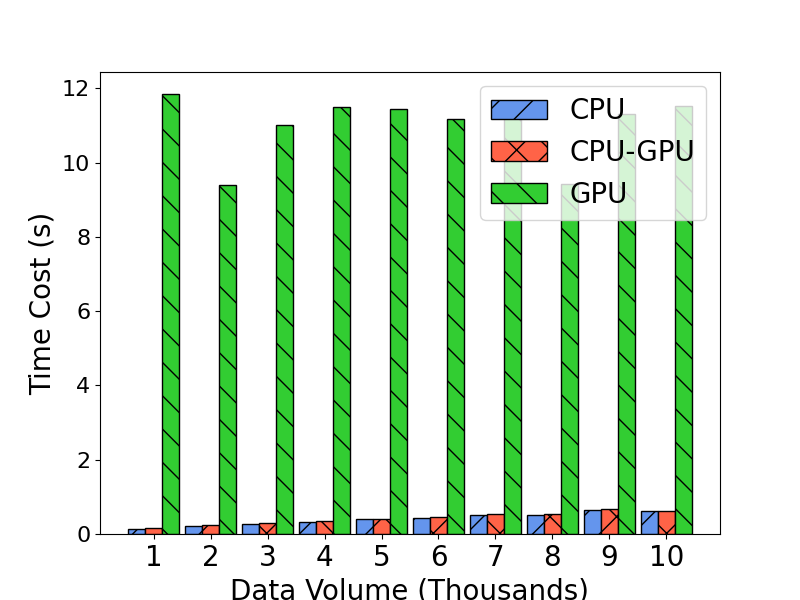}
        \caption{Slice}
        \label{fig:Slice_device}
    \end{subfigure}
    \begin{subfigure}{0.32\linewidth}
        \centering
        \includegraphics[width=\linewidth]{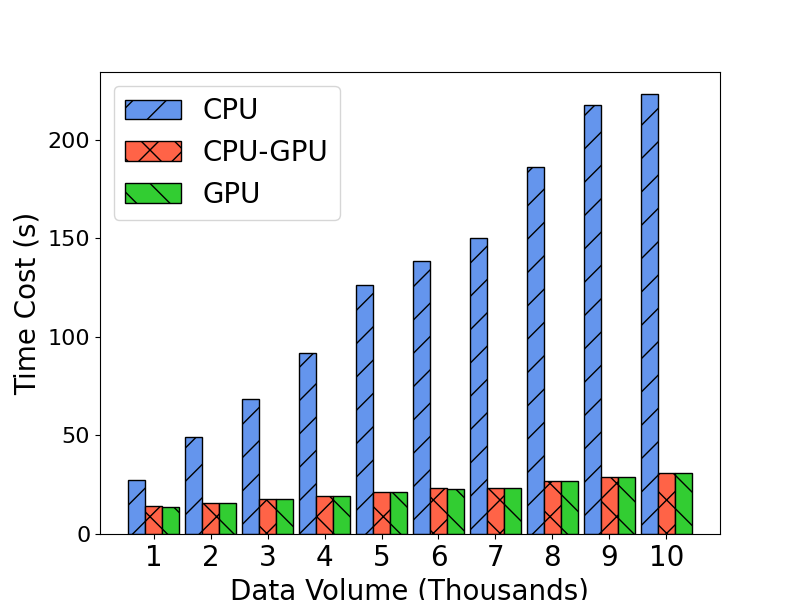}
        \caption{SST-2}
        \label{fig:SST2_device}
    \end{subfigure}    
    \begin{subfigure}{0.32\linewidth}
        \centering
        \includegraphics[width=\linewidth]{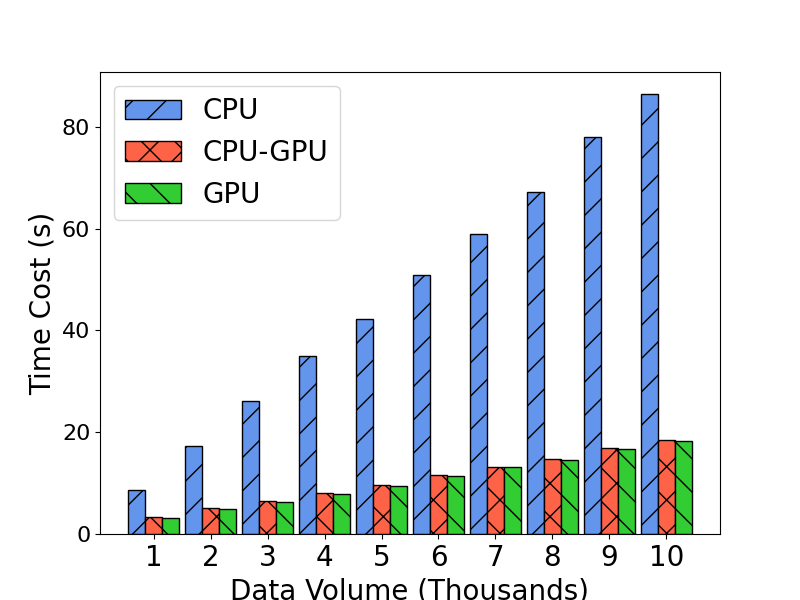}
        \caption{CIFAR-10}
        \label{fig:CIFAR-10_device}
    \end{subfigure}
    \caption{Comparison of different devices}
    \label{fig:gpu_comparison}
\end{figure}

\subsection{Effect of CPU-GPU Placement}
\label{subsec:cpu-gpu}
We evaluate the effectiveness of our cost-based CPU-GPU placement strategy across various dimensions: (1) heterogeneous task types with different operators, cardinality, and data volumes; (2) varying data skew and complex multi-modal queries.

\textbf{Heterogeneous task types.}
We first evaluate inference performance across diverse task types—series, text, and image data—which represent varying model complexities, cardinality, and different operators. 
Figure~\ref{fig:gpu_comparison} shows that CPU outperforms GPU in series tasks due to the minimal computational load and the high overhead introduced by memory-to-GPU data transfer. 
The cost of transferring data to the GPU outweighs the potential gains in computation speed. 
Text and image tasks illustrated in Figures~\ref{fig:SST2_device} and~\ref{fig:CIFAR-10_device} benefit from GPU acceleration. These tasks involve more computationally intensive models, where the GPU's parallelism and higher throughput compensate for the initial data transfer overhead. Our cost model accurately identifies the lower-latency execution device, achieving near-optimal performance with minimal deviation.

\begin{figure}
    \centering
    \begin{subfigure}{0.49\linewidth}
        \centering
        \includegraphics[width=\linewidth]{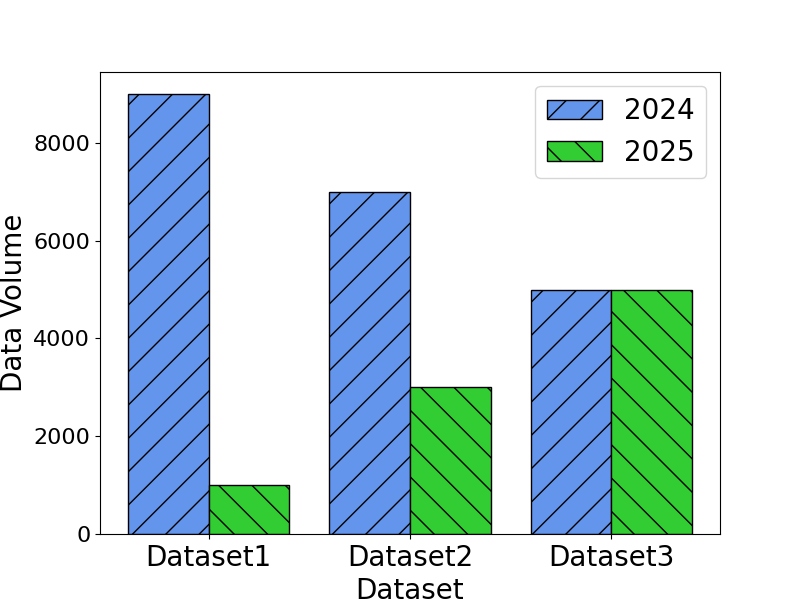}
        \caption{Data Distribution}
        \label{fig:data_distribution}
    \end{subfigure}%
    \hfill%
    \begin{subfigure}{0.49\linewidth}
        \centering
        \includegraphics[width=\linewidth]{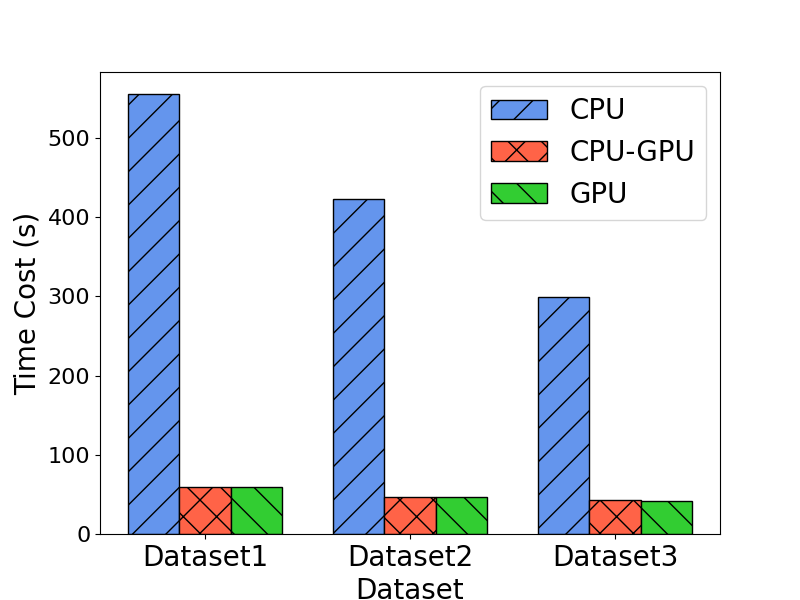}
        \caption{Time Cost}
        \label{fig:data_distribution_cpu_gpu}
    \end{subfigure}
    \caption{Comparison of data skew.}
    \label{fig:different_data_skew}
\end{figure}

\textbf{Data skew.}
We investigate the effectiveness of the cost model under varying degrees of data skew, where input data is unevenly distributed across temporal partitions. As illustrated in Figure~\ref{fig:data_distribution}, we construct three datasets in which records from the year 2024 account for 90\%, 70\%, and 50\% of the data, respectively. SQL queries are issued to filter and perform inference exclusively on 2024 data.
Figure~\ref{fig:data_distribution_cpu_gpu} compares inference performance across CPU, GPU, and CPU-GPU co-processing. 
The results demonstrate that even under skewed distributions, the cost model consistently selects the optimal device, effectively adapting to data locality and operator selectivity to minimize inference latency.

\begin{figure}
    \centering
    \begin{subfigure}{0.49\linewidth}
        \centering
        \includegraphics[width=\linewidth]{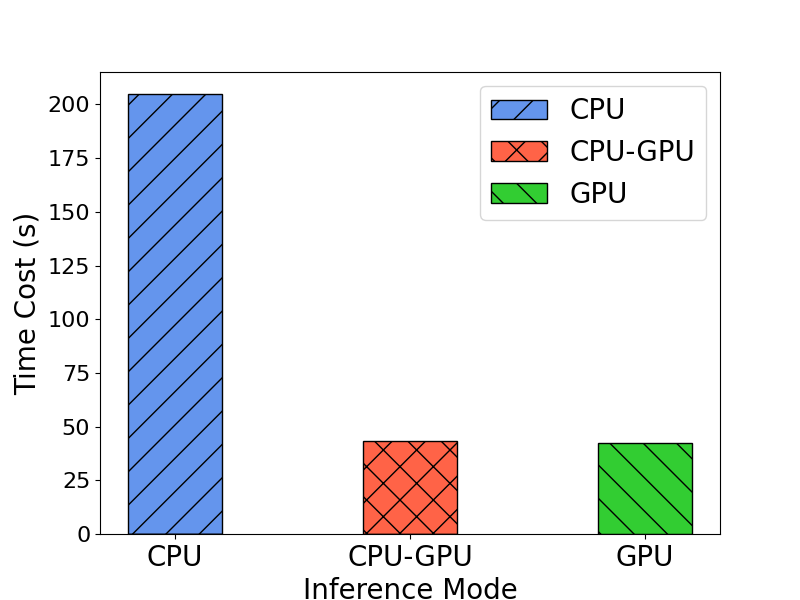}
        \caption{Multi-modal}
        \label{fig:multiple_model}
    \end{subfigure}%
    \hfill%
    \begin{subfigure}{0.49\linewidth}
        \centering
        \includegraphics[width=\linewidth]{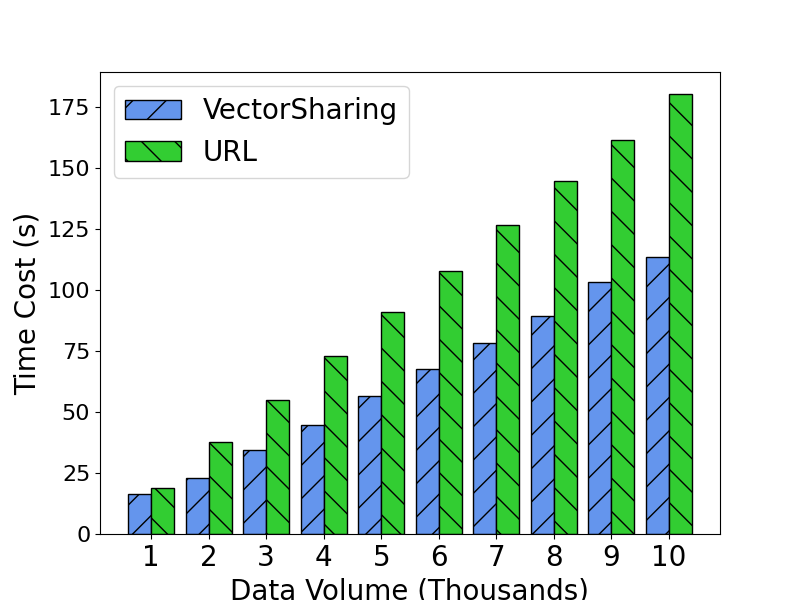}
        \caption{Vector Sharing}
        \label{fig:url_vector}
    \end{subfigure}
    \caption{Multi-modal and Vector Sharing}
    \label{fig:mult-modal_vector_sharing}
\end{figure}

\textbf{Multi-modal.}
We leverage selection and join operators to enable cross-modal data fusion by integrating image recognition and text reasoning models along with window functions applied to the data inputs.
Figure~\ref{fig:multiple_model} shows strong adaptability by assigning heterogeneous devices (e.g., GPU for image models and CPU for lightweight text models) to different sub-tasks. This device-aware execution enables global latency minimization, validating the effectiveness of our model in complex, heterogeneous inference scenarios.

\begin{figure}
    \centering
    \captionsetup[subfigure]{skip=4pt}
    \begin{subfigure}{0.32\linewidth}
        \centering
        \includegraphics[width=\linewidth]{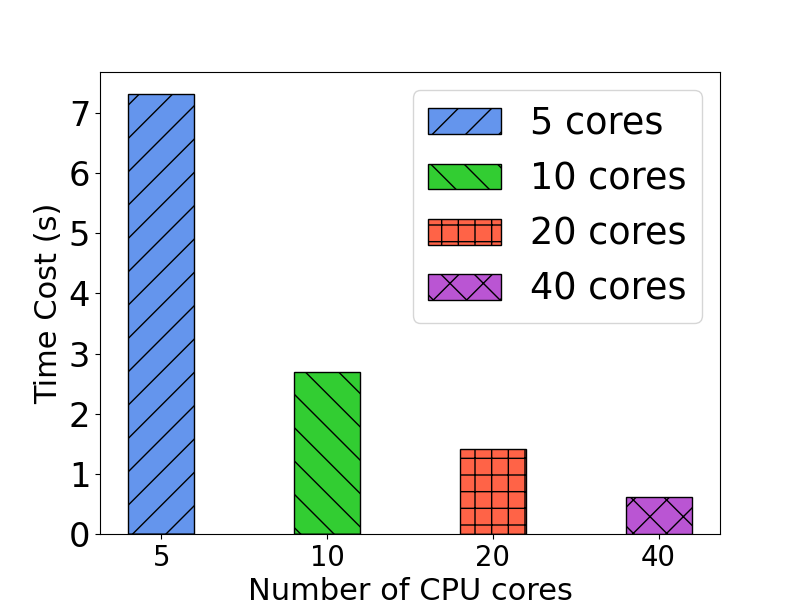}
        \caption{Slice}
        \label{fig:Slice_cpus}
    \end{subfigure}
    \begin{subfigure}{0.32\linewidth}
        \centering
        \includegraphics[width=\linewidth]{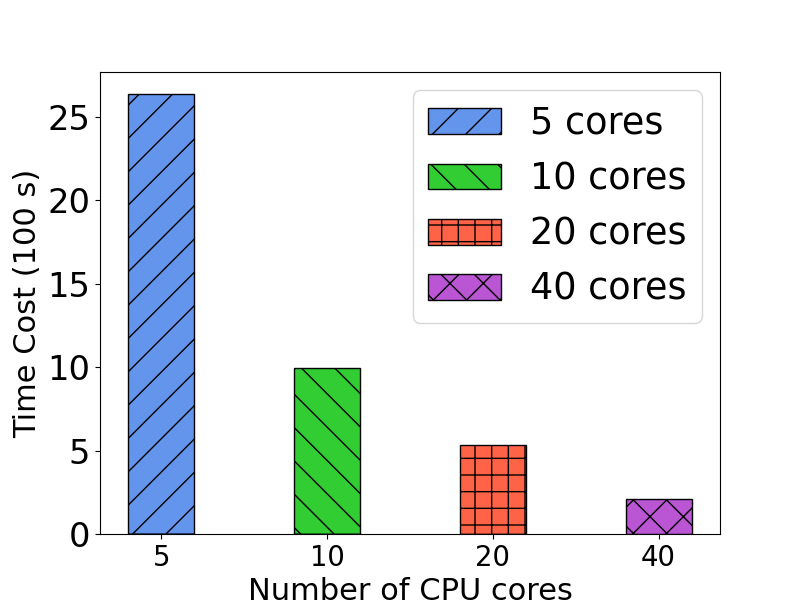}
        \caption{SST-2}
        \label{fig:SST2_cpus}
    \end{subfigure}
    \begin{subfigure}{0.32\linewidth}
        \centering
        \includegraphics[width=\linewidth]{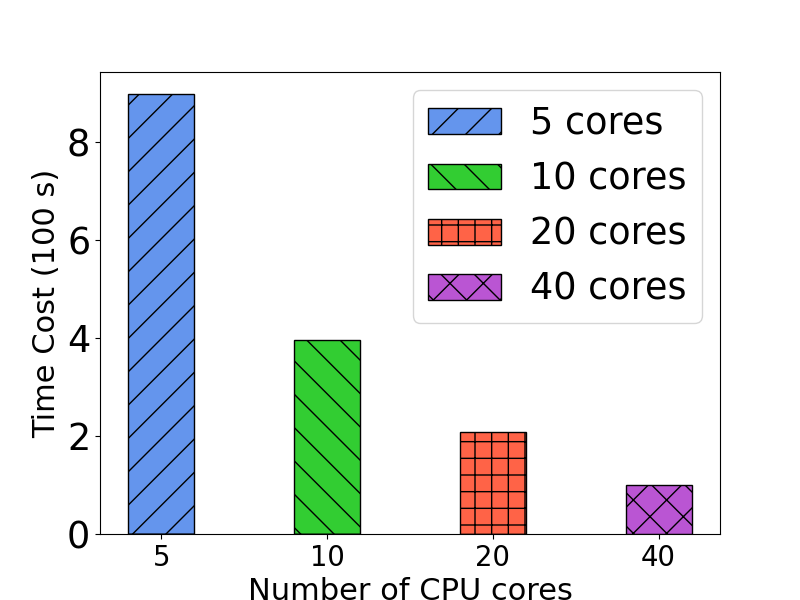}
        \caption{CIFAR-10}
        \label{fig:CIFAR-10_cpus}
    \end{subfigure}    
    \caption{Comparison of CPU configs.}
    \label{fig:cpus_comparison}
\end{figure}

\begin{figure}
    \centering
    \captionsetup[subfigure]{skip=4pt}
    \begin{subfigure}{0.32\linewidth}
        \centering
        \includegraphics[width=\linewidth]{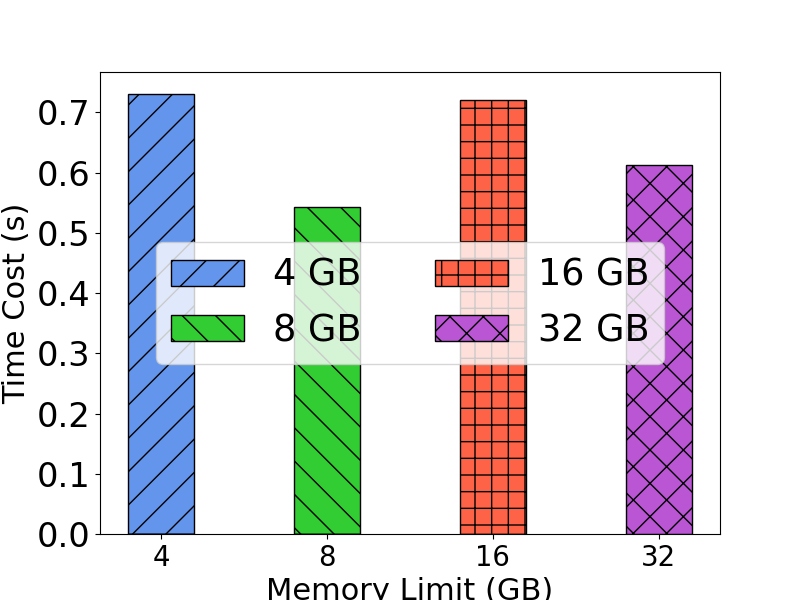}
        \caption{Slice}
        \label{fig:Slice_mems}
    \end{subfigure}
    \begin{subfigure}{0.32\linewidth}
        \centering
        \includegraphics[width=\linewidth]{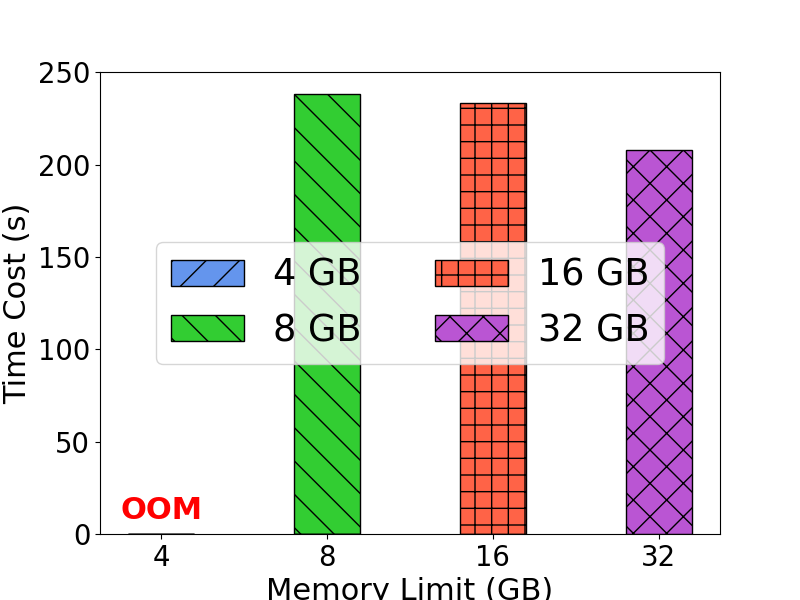}
        \caption{SST-2}
        \label{fig:SST2_mems}
    \end{subfigure}
    \begin{subfigure}{0.32\linewidth}
        \centering
        \includegraphics[width=\linewidth]{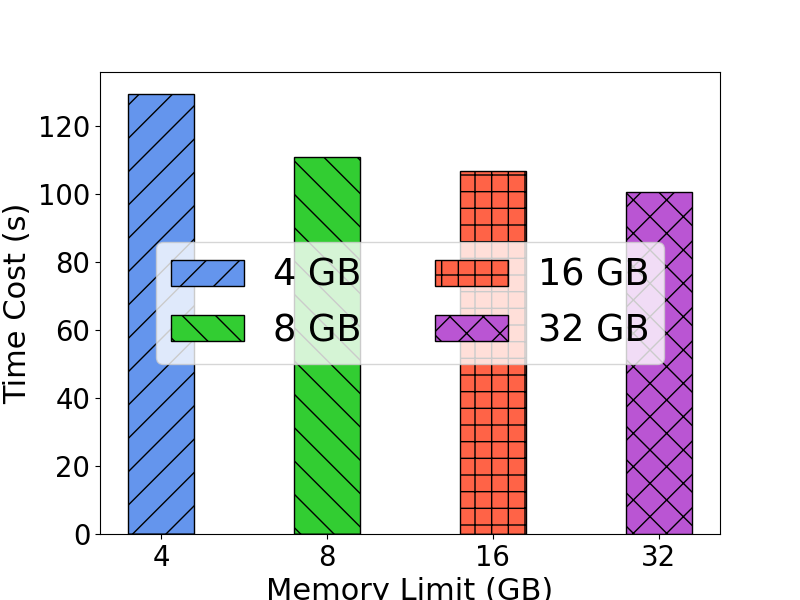}
        \caption{CIFAR-10}
        \label{fig:CIFAR-10_mems}
    \end{subfigure}
    \caption{Comparison of memory configs.}
    \label{fig:mem_comparison}
\end{figure}

\subsection{Ablation study}
\label{subsec:exp_ablation}

In this section, we conduct experiments to validate the advantage of pre-processing vector-sharing storage, pipeline-based batch inference, and resource allocation configurations on {\morphDB} performance.

\subsubsection{Effect of Vector Sharing}

This experiment is designed to assess inference efficiency by comparing two approaches: sharing vector data as pre-processed and embedding vectors within the database versus referencing images by URLs. As shown in Figure ~\ref{fig:url_vector}, sharing vectors directly in the database substantially improves efficiency. 
This advantage stems from the reduced latency associated with in-database vector access, as it eliminates the overhead of external data retrieval and minimizes the delays linked to URL-based storage. By centralizing data within the database, vector storage enhances both retrieval speed and overall inference throughput, making it a superior approach for AI-driven tasks.

\subsubsection{Effect of Batch Size}

We also investigate the performance of various batch sizes for our optimization strategy - batch inference in pipeline.
In our experiments shown in Table \ref{tab:batch}, we evaluate {\morphDB}'s inference performance (time cost) across different batch sizes, including 4, 8, 16, 32, 64, and 128.
The results reveal that there is an optimal batch size range that maximizes computational efficiency and high utilization, beyond which performance gains diminish or even degrade due to resource constraints.
We can find that batch size 8 yields the best performance for the \textit{ImageNet} dataset, while batch size 16 is optimal for the \textit{Finance} dataset.
Choosing the most appropriate batch size is essential to balancing concurrency and resource usage. Smaller batch sizes provide higher concurrency, enabling rapid response times for individual queries but potentially underutilizing computational resources. Conversely, larger batch sizes yield higher throughput by fully utilizing the hardware, but they may limit concurrency.

\begin{table}
  \caption{Effect of batch size}
  \label{tab:batch}
  \begin{tabular}{lcccccc}
    \toprule
    \multirow{2}{*}{Dataset} & \multicolumn{6}{c}{Batch Size (Inference Time in Seconds)} \\
    \cmidrule{2-7}
    & 4 & 8 & 16 & 32 & 64 & 128 \\
    \midrule
    ImageNet & 108.31 & \textbf{81.34} & 86.72 & 84.49 & 91.63 & 106.80 \\
    Finance & 180.35 & 136.86 & \textbf{119.08} & 135.06 & 143.40 & 192.12 \\
    \bottomrule
  \end{tabular}
\end{table}

\subsubsection{Effect of Different resource allocations}

We evaluate task performance by imposing different resource allocation limits. On the GPU server, we restrict the CPU and memory usage of the {\morphDB} Docker image. The experiments are conducted using the same model and dataset across three tasks: the slice dataset for the series task, the Finance dataset for the text task, and the CIFAR-10 dataset for the image task.
As shown in Figure~\ref{fig:cpus_comparison}, after limiting the number of CPUs used to 5, 10, 20, and 40, the model processes 10,000 rows of data across different tasks, and the inference time is significantly reduced, with the reduction being non-linear.
It is worth noting that when limiting memory usage to 4GB, 8GB, 16GB, and 32GB, the high memory demand for the text and image tasks led to out-of-memory (OOM) errors. Specifically, the text task encounters OOM under 4GB memory conditions. Meanwhile, for image and text tasks, the inference time decreases with increasing memory, and for series tasks, the inference time remains roughly the same in Figure~\ref{fig:mem_comparison}.

\section{Related Works}
\label{sec:related_works}

\subsection{In-database AI Analytics} 
\label{subsec:DB_ML}
In-database AI analytics represents a paradigm shift where artificial intelligence (AI) and machine learning (ML) capabilities are directly integrated within the database system, enabling seamless data analysis and model inference without external computation. 
This fusion can be approached through two primary avenues: \textbf{Functional Integration} and \textbf{Kernel-Level Integration}. 

The first one is \textbf{Functional Integration}, focusing on integrating external AI/ML frameworks, such as TensorFlow or PyTorch, or utilizing user-defined functions (UDFs) to execute machine learning tasks in database systems ~\cite{mindsdb, NeurDB, evadb, park2022end, klabe2023exploration, amazonsagemaker, googleaiplatform, ibmdb2}. Users can leverage SQL queries, UDFs, or APIs to invoke these models and perform predictive analytics directly within the database, such as MindsDB~\cite{mindsdb}, EvaDB~\cite{evadb}, and Raven~\cite{karanasos2020extending, park2022end}. 
MindsDB~\cite{mindsdb} is an open-source platform that integrates ML capabilities directly into databases, allowing users to train, test, and deploy models without leaving the database environment.
EvaDB~\cite{evadb} offers a SQL-based interface for building AI applications over both structured and unstructured data, supporting tasks such as regression, classification, image recognition, and generative AI.
Klabe et al.~\cite{klabe2023exploration} explore integrating ML inference into databases via user-defined functions (UDFs) and external ML runtime APIs, enhancing flexibility and interoperability with existing AI tools.
Raven~\cite{karanasos2020extending, park2022end} is a production-ready system that co-locates data and ML runtimes, using a unified graph abstraction for cross-runtime optimization and efficient query execution across heterogeneous hardware.

Kernel-level integration, on the other hand, by incorporating AI and ML algorithms into the core of the database engine, optimizing query processing and computational efficiency~\cite{madlib,li2017mlog, NeurDB, wang2020sqlflow, salazar2024inferdb, salazar2024inferdb, guoliang2024gaussml, oracleautonomous, microsoftazure, saphana, jasny2020db4ml}. 
Madlib~\cite{madlib} facilitates advanced analytics and ML capabilities within a DBMS, enhancing performance and streamlining the analytics workflow. By leveraging distributed computing frameworks like Apache Hadoop and Apache Spark, Madlib enables efficient processing of large datasets without the need to transfer data between systems.  
MLog~\cite{li2017mlog} enables declarative model training and prediction via SQL-like syntax, simplifying the specification and execution of ML workflows within a DBMS.
NeurDB~\cite{NeurDB} integrates AI operators—such as training, inference, and fine-tuning—directly into the DBMS, supporting adaptive analytics under data and workload drift.
InferDB~\cite{salazar2024inferdb} optimizes in-database inference by transforming ML models into index-like structures, enabling low-overhead, index-based predictions.
GaussML~\cite{guoliang2024gaussml} offers a full-stack ML pipeline within the DBMS, covering preprocessing, training, evaluation, and deployment to reduce data movement and improve system efficiency.

{\morphDB} adopts the second integration approach by embedding inference algorithms as native query operators, enabling seamless execution of model inference within the query pipeline. It supports hardware acceleration (e.g., GPUs) and extends the engine with AI-native data structures—such as tensors and vector formats—to ensure flexible model storage and efficient large-scale inference.



\subsection{Automated Machine Learning} 
\label{subsec:AutoML}

Automated Machine Learning (AutoML)~\cite{karmaker2021automl} has emerged as a transformative approach in the field of artificial intelligence, aiming to streamline and automate the end-to-end process of applying machine learning to real-world problems without requiring deep expertise in data science or machine learning.
The core task of AutoML is model selection, which is to train and evaluate several models and automatically identify the most suitable model for a given dataset and task without requiring manual intervention.

AutoSklearn~\cite{autosklearn} partitions the dataset into training and validation sets using the holdout method to estimate validation loss. It employs a bandit-based strategy, specifically successive halving (SH), to efficiently allocate resources across candidate pipelines.
AutoKeras~\cite{autokeras} leverages the evaluation information to define a search space that includes both neural architecture patterns and common hyperparameters. It also tunes hyperparameters from preprocessing steps and the training process to optimize model performance.
AutoGluon~\cite{autogluon} implements a two-stage selection procedure. It first selects the top five models for image, text, and tabular data based on their unimodal benchmarking performance, and subsequently performs a random search over combinations of these models along with additional hyperparameters. The optimal combination identified through this process serves as the default configuration.
TRAILS~\cite{xing2024database} proposes an SLO-aware and resource-efficient two-phase model selection algorithm with a novel coordinator to leverage the strengths of both efficient training-free and effective training-based model selection.


In contemporary AutoML frameworks, model selection focuses on the evaluation metrics of neural architectures and hyperparameter selection ~\cite{he2021automl}, followed by full training on the target task which is computationally intensive. 
In contrast, our model selection approach aligns more closely with transfer learning~\cite{zhuang2020transfer_learning_survey}, where the most suitable pre-trained model is selected for the target task. Compared to training from scratch, this paradigm not only leverages the rich knowledge learned from the source task but also significantly accelerates the training process.

\section{Conclusion}
\label{sec:conclusion}


In this paper, we present {\morphDB}, a task‑centric AI‑native DBMS implemented as a PostgreSQL extension with LibTorch, designed to automate model management, selection, and inference within the database engine.
{\morphDB} introduces specialized schemas and data types with an Mvec representation for 3D tensors, supporting both BLOB‑based all‑in‑one and decoupled architecture–parameter formats.
Model selection in {\morphDB} is formulated as a transfer learning problem, addressed through a two-phase framework. An offline embedding phase constructs a transferability subspace from historical tasks, and an online projection phase maps new tasks into this subspace to identify suitable models with minimal overhead..
Furthermore, two optimization strategies: pre‑embedding with vectoring sharing and DAG‑based batch pipelines,  are proposed to improve the inference throughput. 
Extensive evaluation on nine public datasets spanning time-series forecasting, natural language processing, and image recognition demonstrates that {\morphDB} achieves superior performance compared to existing AI-native DBMSs (e.g., EvaDB, MADlib, GaussML) and state-of-the-art AutoML frameworks (e.g., AutoGluon, AutoKeras, AutoSklearn), offering a robust trade-off between prediction accuracy, resource efficiency, and model selection latency.

\begin{acks}
This work was supported by the Key Research Program of Zhejiang Province (No. 2023C01037). It was also supported by the National Natural Science Foundation of China (NSFC) under Grant No. 62572422.
\end{acks}

\bibliographystyle{ACM-Reference-Format}
\bibliography{main}





\end{document}